\documentclass[preprint]{elsarticle}

\newcommand{\refeq}[1]{Eq. (\ref{#1})}
\newcommand{\refeqs}[2]{Eq. (\ref{#1}$-$\ref{#2})}
\newcommand{\reffig}[1]{figure \ref{#1}}
\newcommand{\Reffig}[1]{Figure \ref{#1}}

\newcommand{\refsec}[1]{Section \ref{#1}}

\usepackage{bm}
\usepackage{amsmath}
\usepackage[font=footnotesize,labelfont=bf]{caption}
\usepackage[font=footnotesize,labelfont=bf]{subcaption}
\usepackage{algorithm}
\usepackage{algpseudocode}
\usepackage[hyphens]{url}
\usepackage{hyperref}
\hypersetup{breaklinks=true}

\begin{document}

\title{A Cartesian-octree adaptive front-tracking solver for immersed biological capsules in large complex domains}
\author[1]{Damien P. Huet}
\ead{huet@math.ubc.ca}
\author[1,2]{Anthony Wachs\corref{cor1}}
\ead{wachs@math.ubc.ca}
\cortext[cor1]{Corresponding author}
\affiliation[1]{organization={Department of Mathematics, University of British Columbia}, addressline={1984 Mathematics Road}, postcode={BC V6T 1Z2}, city={Vancouver}, country={Canada}}
\affiliation[2]{organization={Department of Chemical \& Biological Engineering, University of British Columbia}, addressline={2360 E Mall}, postcode={BC V6T 1Z3}, city={Vancouver}, country={Canada}}

\begin{abstract}
  We present an open-source adaptive front-tracking solver for biological capsules in viscous flows. The membrane elastic and bending forces are solved on a Lagrangian triangulation using a linear Finite Element Method and a paraboloid fitting method. The fluid flow is solved on an octree adaptive grid using the open-source platform Basilisk. The Lagrangian and Eulerian grids communicate using an Immersed Boundary Method by means of Peskin-like regularized Dirac delta functions.
  We demonstrate the accuracy of our solver with extensive validations: in Stokes conditions against the Boundary Integral Method, and in the presence of inertia against similar (but not adaptive) front-tracking solvers. Excellent qualitative and quantitative agreements are shown. We then demonstrate the robustness of the present solver in a challenging case of extreme membrane deformation, and illustrate its capability to simulate inertial capsule-laden flows in complex STL-defined geometries, opening the door for bioengineering applications featuring large three-dimensional channel structures. The source code and all the test cases presented in this paper are freely available.
\end{abstract}

\maketitle

\section{Introduction}
The numerical study of membrane-enclosed fluid objects, or capsules, has seen tremendous interest over the past three decades due to the wide range of applications in the biomedical and bioengineering world. Indeed, numerical simulation of capsule dynamics in viscous flows is crucial to better characterize and understand blood flow through capillary microcirculation and develop applications such as targeted drug delivery \cite{dewhirst2017transport}, migration of cancerous leukocytes through the microvascular network \cite{puleri2021computational, balogh2021data} and cell sorting and cell characterization in microfluidic devices \cite{islamzada2020deformability, bazaz2022zigzag}.
In particular, the latter application has the potential to speed-up labour-intensive diagnosis procedures or to extract relevant components of biofluids. For instance intertial centrifugation in spiral-shaped microchannels has been shown to efficiently and accurately segregate cells based on their size, and could be applied to perform non-destructive blood plasma extraction \cite{takeishi2022inertial, gangadhar2022inertial, fang2022efficient}.

The study of capsules from a mechanical point of view was paved in 1981 by the pioneering analytical work of Barthès-Biesel \& Rallison \cite{barthes1981time}, who derived from the thin-shell theory a time-dependant expression for the deformation of an elastic capsule in a shear flow in the limit of small deformations.
A decade later, Pozrikidis went beyond the assumption of small deformations, using the Boundary Integral Method (BIM) to investigate finite deformations of elastic capsules in a shear flow \cite{pozrikidis1995finite, ramanujan1998deformation}. This work was quickly followed by Eggleton \& Popel who simulated spherical and biconcave capsules in shear flows using the Front-Tracking Method (FTM) \cite{eggleton1998large}.
Capitalizing on the advantages of the BIM $-$ such as a lower computing cost compared to the FTM, and the ability to simulate true Stokes conditions $-$ Pozrikidis investigated the bending resistance of capsules and proposed a simplified bending model for biological membranes valid for small deformations \cite{pozrikidis2001effect}, leading to the first numerical simulation of an RBC based on the thin-shell theory \cite{pozrikidis2003numerical}.
The work of Pozrikidis was later extended by Zhao et al., who proposed a BIM able to simulate RBCs in complex geometries with up to 30\% volume fraction \cite{zhao2010spectral}. In the 2000s, Barthès-Biesel and Lac also used the BIM and studied finite deformations of capsules devoid of bending resistance: they considered the effect of the membrane constitutive law and exhibited buckling instabilities \cite{lac2004spherical} as well as the dynamics of two interacting capsules in a shear flow \cite{lac2007hydrodynamic}.

Despite the major success of the BIM to simulate capsules and biological cells, the FTM is still being developed. Indeed, while the FTM is more computationally intensive than the BIM because it necessitates meshing the whole 3D fluid domain, and while it can require very small time steps to satisfy stability conditions depending on the considered membrane forces; the FTM can handle inertial regimes, thus allowing to examine a wider range of applications.
As such, Bagchi uses the FTM to perform two-dimensional simulations of several thousand RBCs in a shear flow \cite{bagchi2007mesoscale}, allowing the study of RBC interactions at the mesoscale.
In the next years, Doddi \& Bagchi \cite{doddi2008lateral} and Yazdani \& Bagchi \cite{yazdani2011phase, yazdani2012three, yazdani2013influence} develop respectively three-dimensional implementations of the elastic membrane stress and of the Helfrich's bending stress for biological membranes, the latter not being limited to small deformations as was the case for the formulation of Pozrikidis. The ability to consider finite Reynolds numbers allowed Doddi \& Bagchi to extend the work of Lac et al. on capsule interactions to intertial flows \cite{doddi2008effect}. Their framework was later extended to complex geometries by Balogh \& Bagchi \cite{balogh2017computational}, enabling them to study the dynamics of hundreds of RBCs in a microvascular network with a hematocrit (volume fraction of RBCs) of 30\% \cite{balogh2018analysis}.
Another variant of the FTM is to use a Lattice-Bolztman fluid solver rather than the traditional PDE-based Navier-Stokes solver: this can bring significant performance improvement especially at low Reynolds number where the Lattice-Boltzman Method (LBM) performs well. For instance, Li \& Sarkar extend the work of Barthès-Biesel and Lac on the instabilities of elastic capsules in shear flows using an FTM-LBM solver \cite{li2008front}, and Zhang et al. describe a similar framework able to simulate RBCs \cite{zhang2007immersed}, including cell-cell aggregation phenomena \cite{zhang2008red}. More recently, Ames et al. \cite{ames2020multi} harnessed the performance improvements of GPUs and demonstrated an impressive 17 million RBCs simulated in a microvascular network using a similar FTM-LBM framework.

Other methods to simulate biological capsules and vesicles include the RBC model of Fedosov et al. \cite{fedosov2010multiscale}. In the work of Fedosov, the RBC membrane mechanics is not governed by the thin shell theory but rather by a coarse-grained molecular dynamics model. The membrane of the RBC is discretized, with each edge representing several nonlinear springs which correspond to elastic and viscoelastic properties of the membrane of the RBC. The model parameters are found for an extremely fine mesh of over $27000$ nodes, where the lengths of the edges correspond to that of biological spectrins. Yet, in Fedosov's model practical RBC simulations are conducted with different model parameters which are intended to display the same mechanical behavior as that obtained with a fine mesh, but using a number of nodes orders of magnitude lower. These coarse-grained mechanical properties of the membrane are shown to lead to results which accuracy lies within the range of experimental measurement errors for the specific cases considered. However the range of validity of this coarse-graining step is not obvious and the spatial convergence can be non-monotonous or even not exist (see the transverse diameter plot in figure 1 in \cite{fedosov2010multiscale}), indicating to use this model with care. A last approach to capsule simulations is to adopt a fully Eulerian framework, where the membrane is not discretized with Lagrangian nodes, edges and faces. Instead, the capsule configuration is described using the Eulerian grid employed to solve the Navier-Stokes equations. Removing the need of a Lagrangian grid is a desirable description as the IBM can reduce the spatial and temporal accuracies to first order if no special treatment is implemented. In the Eulerian capsule description, Volume-Of-Fluid (VOF), level-set or phase-field methods can be utilized to track the position of the membrane, similar to what is done in the context of fluid-fluid interfaces \cite{hirt1981volume, brackbill1992continuum, tryggvason2001front, popinet2009accurate}.
If the considered membrane mechanical behavior is independant of the past configuration of the membrane, for instance if there is no resistance to shear and high resistance to bending, the membrane forces can be computed using techniques developed for surface-tension flows: the local curvature can be computed using height-functions in the case of a VOF description, or by numerically differentiating the level-set or phase-field function near the interface \cite{cottet2006level}. However, in most biological applications the membrane properties do depend on the past membrane configurations due to its elastic behvior. In such cases a quantity representing the membrane stretch needs to be initialized and advected in the vicinity of the membrane, for instance the left Cauchy-Green deformation tensor. Ii et al. have demonstrated that this approach is possible and scalable \cite{ii2012full,ii2012computational,ii2018continuum}, although more comparisons with the FTM are needed in order to evalute the performance and the accuracy of the Eulerian methods, especially for long-lasting simulation.

Concomitent to these developments of capsule simulations, the IBM gained great popularity in the particle-laden flows community \cite{Uhlmann2005, Kempe2012a, breugem2012second}, and in the past two decades some adaptive IBM have been proposed in cases of immersed solid particles. In this context, Roma et al. \cite{Roma1999} present a two-dimensional adaptive IBM implementation where the Adaptive Mesh Refinement (AMR) is achieved by means of overlapping rectangles $-$ or ``patches" $-$ of finer grid cells in the regions of interest, i.e. where higher accuracy of the flow field is needed. This method was later improved and proved second-order accurate by Griffith et al. \cite{griffith2007adaptive} and Vanella et al. \cite{vanella2014adaptive}.
Previously, Agreasar et al. \cite{agresar1998adaptive} had used a non-patched adaptive FTM-IBM method in order to simulate axisymmetric circular cells. Their IBM implementation did not use Peskin-like regularized Dirac delta functions: instead the Lagrangian grid on the membrane communicates with the background Eulerian grid via an area-weighted extrapolation.  More recently, Cheng \& Wachs \cite{cheng2022immersed} used the IBM coupled with an LBM solver to achieve adaptive simulations in the case of a single rigid sphere in various flow conditions.

The goal of this paper is to present an efficient framework to study the dynamics of dilute suspensions of capsules in complex geometries, not limited to non-inertial regimes. As the BIM cannot be used at finite Reynolds numbers, we use the FTM and therefore the whole 3D fluid domain is discretized.
Since a vast range of realistic applications consider geometries of sizes orders of magnitude larger than the typical size of a capsule, requiring hundreds of millions to billions of Eulerian grid cells when the Cartesian grid has a constant grid size, we develop an adaptive FTM solver rendering achievable to simulate configurations that were previously out of reach with a constant grid size.
We provide the open-source code as part of the Basilisk platform \cite{popinet2003gerris, popinet2009accurate, Popinet2015, huet_sandbox}.

The paper is organized as follows: in \refsec{sec:governing_eq} we present the problem formulation and the governing equations for both the fluid and the capsule dynamics. We describe the implementation of our numerical model in \refsec{sec:implementation}, emphasizing the finite element membrane model and the FTM method. Numerous validation cases are shown in \refsec{sec:validation}, for increasingly difficult configurations: validations are performed by comparing our computed results against accurate BIM data available in literature whenever possible, otherwise against other FTM results. \refsec{sec:results} contains new results generated with the present method, where the adaptive mesh capability dramatically improves computational efficiency. In \refsec{sec:conclusion} we summarize our work and discuss the strengths, weaknesses and possible improvements of the present method as well as future perspectives.

\section{Governing equations \label{sec:governing_eq}}
\subsection{Fluid motion}

The fluid phase is assumed Newtonian and incompressible: the fluid surrounding and enclosed by the elastic membranes is described using the mixture Navier-Stokes equations:
\begin{equation}
  \label{eq:ns_momentum}
  \rho\left(\frac{\partial \bm{u}}{\partial t} + \bm{u} \cdot \nabla \bm{u}\right) = -\nabla p + \nabla \cdot \left( \mu \left( \nabla \bm{u} + (\nabla \bm{u})^T \right) \right) + \bm{f_b}
\end{equation}
\begin{equation}
  \label{eq:ns_mass}
  \nabla \cdot \bm{u} = 0
\end{equation}
where $\bm{u}$ is the velocity, $p$ is the pressure, $\rho$ is the constant density and $\mu$ is the variable viscosity field, since we will consider non-unity viscosity ratios $\lambda_\mu = \mu_i / \mu_e \neq 1$, with $\mu_i$ and $\mu_e$ the internal and external viscosities. $\bm{f_b}$ denotes the body force containing the membrane elastic and bending force densities acting on the fluid: $\bm{f_b} = \bm{f}_\text{elastic} + \bm{f}_\text{bending} = \left(  \bm{F}_\text{elastic} + \bm{F}_\text{bending} \right) / V$, with $V$ a relevant control volume and $\bm{F}_\text{elastic}$ and $\bm{F}_\text{bending}$ the integrated membrane force densities.

\subsection{Membrane mechanics}
We assume that the lipid-bilayer membrane is infinitely thin: please note that this is not a strong assumption for most biological cells, as thickness of the biological membrane (lipid-bilayer) is $5nm$ while an RBC characteristic size is $10\mu m$. A biological membrane undergoing deformation responds with elastic and bending stresses, described with two distinct mechanical models.

The elastic strains and stresses are described using the theory of thin shells\cite{green1960large}. We summarize this framework here, but the interested reader is referred to the work of \cite{barthes1981time} for more details.
In this continuous description of the capsule, we first introduce the projectors $\bm{P} = \bm{I} - \bm{n} \bm{n}$ and $\bm{P_R} = \bm{I} - \bm{n_R} \bm{n_R}$ onto the current and reference (stress-free) membranes shapes, with $\bm{I}$ the identity tensor and $\bm{n}$ and $\bm{n_R}$ the unit normal vectors to the current and reference membranes configurations, which are both oriented outward.
The membrane strains are described using the surface deformation gradient tensor $\bm{F_s}$, derived from the classical deformation gradient tensor $\bm{F}$ as follows:
\begin{equation}
  \bm{F_s} = \bm{P} \cdot \bm{F} \cdot \bm{P_R}.
  \label{eq:fs}
\end{equation}
The surface right Cauchy-Green deformation tensor $C_s$ is then defined from $\bm{F_s}$:
\begin{equation}
  \bm{C_s} = \bm{F_s^T} \cdot \bm{F_s}.
  \label{eq:cs}
\end{equation}
Let the three eigenvalues of $\bm{F_s}$ be $\lambda_1, \lambda_2, 0$ associated with the eigenvectors $\bm{t_1}, \bm{t_2}, \bm{n}$. Then the eigenbasis of $\bm{C_s}$ is the same as that of $\bm{F_s}$, associated with eigenvalues $\lambda_1^2, \lambda_2^2, 0$.
Note that in the stress-free configuration, at the beginning of a typical simulation, $\lambda_1 = \lambda_2 = 1$, and $\bm{F_s} = \bm{C_s} = \bm{P} = \bm{P_R}$.

The above quantities are useful to compute the membrane elastic stress, which can be expressed using a surface strain-energy function $W_s(\lambda_1, \lambda_2)$:
\begin{equation}
  \sigma_i = \frac{1}{\lambda_j}\frac{\partial W_s}{\partial \lambda_i}, \qquad i \neq j.
\end{equation}

In this work, two distinct strain-energy functions corresponding to two membrane elastic laws are used to describe several types of lipid bilayers in various conditions:
\begin{enumerate}
  \item The Neo-Hookean law, used to describe vesicles and artifical capsules, and which corresponding strain-energy function is \begin{equation}
    W_s^{NH} = \frac{E_s}{6} \left( \lambda_1^2 + \lambda_2^2 + \frac{1}{\lambda_1^2 \lambda_2^2} - 3 \right),
  \end{equation}
  where $E_s$ denotes the shear modulus.
  \item The Skalak law, used to describe the elastic response of RBC membranes, and which strain energy function is \begin{equation}
    W_s^{Sk} = \frac{E_s}{4} \left( I_1^2 + 2I_1 - 2I_2 + CI_2^2 \right),
  \end{equation}
  where the invariants $I_1 = \lambda_1^2 + \lambda_2^2 -2$ and $I_2 = \lambda_1^2 \lambda_2^2 - 1$ have been introduced, as well as the area dilatation modulus $C$ preventing strong area changes and is taken ``large" \cite{pozrikidis1995finite, barthes2002effect} in order to describe the strong area incompressibility of RBCs. Unless otherwise stated, the value $C = 10$ is used in the simulation results presented below.
\end{enumerate}
Once the elastic stress is known, the elastic force exterted by the membrane onto the fluid is simply
\begin{equation}
  \bm{F}_\text{elastic} = \nabla \cdot \bm{\sigma},
  \label{eq:f_el}
\end{equation}
although we will follow the approach of \cite{charrier1989free} and use the principle of virtual work instead of directly computing the divergence of the stress, as explained in \refsec{sec:fem}.

The bending stresses are described using Helfrich's bending energy per unit area $\mathcal{E}_B$ \cite{helfrich1973elastic}:
\begin{equation}
  \mathcal{E}_B = 2 E_b \left( \kappa - \kappa_0 \right)^2 + E_g \kappa_g
\end{equation}
where $\kappa = (\kappa_1 + \kappa_2)/2$ is the local mean curvature, $\kappa_g = \kappa_1 \kappa_2$ is the Gaussian curvature, and $E_b$ and $E_g$ are their associated bending modulii. $\kappa_1$ and $\kappa_2$ are the two principal curvatures at a given point of the two-dimensional membrane sheet, and $\kappa_0$ is the reference curvature.
Then, the bending stresses are derived from the total bending energy  $\int_\Gamma \left( 2 E_b \left( \kappa - \kappa_0 \right)^2 + E_g \kappa_g \right) dS $ by means of a variational derivative, to yield the normal bending force per unit area \cite{guckenberger2017theory}:
\begin{equation}
  \bm{F}_\text{bending}/A = -2 E_b ( \Delta_s(\kappa - \kappa_0) + 2 (\kappa - \kappa_0)( \kappa^2 - \kappa_g + \kappa_0\kappa) ) \bm{n},
  \label{eq:bending-force}
\end{equation}
where the operator $\Delta_s$ is the surface Laplacian $-$ or Laplace-Beltrami operator $-$ defined as $\Delta_s = \nabla_s \cdot \nabla_s = \left((\bm{I} - \bm{n}\bm{n}) \cdot \nabla \right) \cdot \left((\bm{I} - \bm{n}\bm{n}) \cdot \nabla \right)$, and $A$ is a relevant control area. Note how $E_g$ has disappeared in the variational formulation because $\kappa_g$ is a topoligical invariant \cite{barthes2016motion, guckenberger2017theory}.

At this point a parallel with surface tension forces is enlightening: the bending energy related to surface tension acting on a droplet is proportional to the area of the interface, and leads to surface tension forces proportional to the curvature, i.e. to the second derivative of the interface geometry. In contrast, as stated above the bending energy related to lipid-bilayer membranes is proportional to the curvature, and thus the corresponding bending force depends on the second derivative of the curvature, i.e. to fourth-order derivatives of the geometry. As such, the numerical simulations of biological capsules subject to bending stresses is a formidable challenge, and the interested reader is referred to the reviews of \cite{guckenberger2016bending, guckenberger2017theory}. Our approach to computing the bending force is described in \refsec{sec:bending-implementation}, and corresponds to method E in \cite{guckenberger2016bending}.

\section{Numerical method \label{sec:implementation}}
\subsection{Adaptive Finite Volume solver for the Navier-Stokes equations}
Assuming the body force field $\bm{f_b}$ is known, \refeqs{eq:ns_momentum}{eq:ns_mass} are solved using the open-source platform Basilisk \cite{Popinet2015}. The viscous term $\nabla \cdot (\mu (\nabla u + (\nabla u)^T))$ is treated implicitely using a multigrid Poisson solver \cite{popinet2009accurate,popinet2003gerris}, the incompressibility condition is satisfied by the classical projection method of Chorin \cite{chorin1968numerical}, and the advection term $\bm{u} \cdot \nabla \bm{u}$ is solved using the second-order Bell-Colella-Glaz upwind advection scheme \cite{bell1989second}.
To this end, the divergence-free velocity field and the viscosity field are located on cell faces while a velocity field approximately divergence-free, the pressure field, and the body force field are all defined on the cell centers.

Basilisk computes the solution of the Navier-Stokes equations on an octree grid, which allows to coarsen and refine computational cells throughout the simulations while keeping a structured mesh.
The coarsening and refinement of grid cells is  implemented using a wavelet-based algoritm: for the sake of completeness we present here a short overview of the adaptivity process, and the interested reader is referred to \cite{popinet2003gerris, popinet2009accurate, van2018towards} for more in-depth descriptions. Let $f$ be a field of interest which variations in space will govern the size of the grid cells. First, $f$ is downsampled onto a lower level grid by volume-averaging: we call this downsampled field $f_d$.
Then, $f_d$ is upsampled back to the original grid using second-order interpolations, resulting in the downsampled-then-upsampled field $f_{du}$. Since $f$ and $f_{du}$ are defined on the same (original) grid, the sampling error field $\epsilon^i = \| f^i - f^i_{du} \|$ can be defined in each computational cell $i$. Finally, $\epsilon^i$ is used to decide if cell $i$ should be coarsened or refined based on an adaptivity criterion $\zeta > 0$:

\begin{equation}
  \begin{cases}
    \epsilon^i > \zeta \quad \Rightarrow \quad \text{refine cell} \; i\\
    \epsilon^i < \frac{2\zeta}{3} \quad \Rightarrow \quad \text{coarsen cell} \; i\\
    \frac{2\zeta}{3} \leq \epsilon^i \leq \zeta \quad \Rightarrow \quad \text{leave cell $i$ at its current level}.
  \end{cases}
\end{equation}

This wavelet-based algorithm is very versatile as any field of interest can be used to influence the refinement level of the octree grid. In this study, the adaptivity is based on the velocity field and on both the presence of domain boundaries and the capsule membrane. In other words, the fields of interest that the wavelet adaptivity algorithm considers are: $u_x$, $u_y$, $u_z$, $c_s$ and $\xi$, where the first three scalar fields are the three components of the velocity field $\bm{u}$, $c_s$ is the fluid volume fraction field (in case of complex geometries), and $\xi$ is a scalar field which varies strongly in the vicinity of the membrane and is constant elsewhere $-$ see \refsec{sec:FTM-adapt} for a definition of $\xi$. The refinement criterion $\zeta$ can be different for different fields of interest.
In this study, we choose $\zeta$ to be very small when applied to $c_s$ and $\xi$ in order to impose the maximum level of refinement in the vicinity of walls and of the membrane (typically we choose $\zeta < 10^{-10}$); while we found by trial and error that having $\zeta$ of the order of $1\%$ of the characteristic velocity when applied to $u_x$, $u_y$ and $u_z$ leads to satisfactory refinement and coarsening of computational cells in the rest of the fluid domain.

\subsection{Second-order treatment of solid embedded boundaries}
\begin{figure}
  \centering
  \includegraphics[height=.3\textheight]{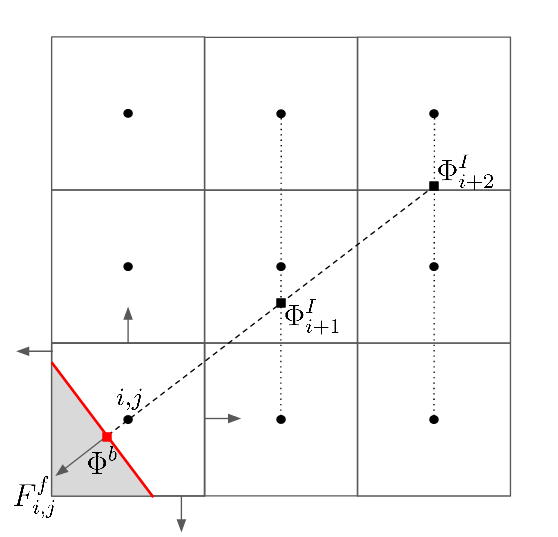}
  \caption{A two-dimensional example of an interpolation stencil estimating the boundary flux $\Phi^b$ with second-order accuracy. The gray area denotes the solid, the red line denotes the solid boundary, and the arrows denote the five face fluxes that are computed in the cut cell of interest $i,j$. The circle dots denote the Eulerian cell centers while the square dots show the locations of the data points $\Phi$ from which the boundary flux $F^f_{i,j}$ is interpolated. The interpolation line normal to the solid boundary is represented by a dashed line. In case of Dirichlet boundary conditions, $\Phi^b$ is known but the data points $\Phi^I_{i+1}$ and $\Phi^I_{i+2}$ are themselves interpolated with second-order accuracy along the dotted lines using the centers of the Eulerian grid cells.}
  \label{fig:cut-cell}
\end{figure}

In Basilisk, complex geometries are handled using the sharp, second-order and conservative embedded boundaries method of Johansen and Colella \cite{johansen1998}. In this method, the solid boundaries are assumed to cut the Eulerian cells in a piecewise linear fashion. Boundary conditions are enforced by estimating the flux on the solid boudary using second-order interpolations on a stencil involving the surrounding fluid cells. An example of such interpolation stencil in two dimensions is shown in \reffig{fig:cut-cell}. This method can be implemented so that only the volume and face fluid fractions are necessary to describe the boundary and recover the boundary flux \cite{popinet2003gerris}. As such, at the beginning of the simulation these two fluid fraction fields are generated from either a user-defined level-set function describing the geometry  or an STL file.

If no additional treatment is done, it is well known that this class of cut-cell methods suffers from unreasonably strict CFL restrictions due to cells with small fluid volume fractions.
Indeed, when a cell is cut by a solid boundary the effective CFL condition becomes $\Delta t < c\Delta x/(f |\bm{u}|)$, where $c$ and $f$ are the volume and face fractions, and $|\bm{u}|$ is the velocity norm in the considered cell. If $c/f$ is close to zero, the time step $\Delta t$ may become arbitrarily small. To alleviate this issue, a ``flux redistribution" technique is carried out, where the fluxes of problematic small cells are redistributed to their neighbors, thus preventing $\Delta t$ from becoming arbitrarily small \cite{colella2006cartesian, embed-small-cfl, ghigo-embed}.

\subsection{Front-Tracking Method (FTM)}
\subsubsection{Standard FTM formulation}

The capsule configuration is described using the Front Tracking Method: we adopt a Lagrangian representation of the capsule, which we cover by a triangulated, unstructured mesh \cite{unverdi1992front, tryggvason2001front}. This Langrangian mesh communicates with the Eulerian octree grid used to decribe the background fluid by means of regularized Dirac-delta functions introduced by Peskin \cite{Peskin1977}, which role is to interpolate velocities from the Eulerian grid to a Lagrangian node; and to spread membrane forces from a Lagrangian node to the Eulerian grid. In this paper we use a cosine-shaped regularized delta function:
\begin{equation}
  \label{eq:dirac-delta}
  \delta(\bm{x_0} - \bm{x}) =
  \begin{cases}
  \begin{aligned}
    & \frac{1}{64 \Delta^3} \prod_{i = 1}^3 \cos\left( \frac{\pi}{2\Delta} (x_{0, i} - x_i) \right) \qquad \text{if} \quad |x_{0,i} - x_i| < 2 \Delta\\
    & 0 \qquad \text{otherwise}
  \end{aligned}
\end{cases},
\end{equation}
where $\Delta$ is the length of an Eulerian cell and $\bm{x_0} = [x_0, y_0, z_0]$ corresponds in practice to the coordinates of a Lagrangian node. The prefactor $1/(64 \Delta^3)$ ensures that the discrete integral over the whole space $\int_\Omega \delta(\bm{x_0} - \bm{x}) d\bm{x}$ is equal to $1$.
Then, the velocity $\bm{u_0}$ of a given Lagrangian node located at $\bm{x_0}$ is interpolated from the Eulerian velocity field $\bm{u}$ using:
\begin{equation}
  \label{eq:eul2lag}
  \bm{u_{0}} = \int_\Omega \bm{u}(\bm{x}) \delta(\bm{x_0} - \bm{x}) d \bm{x} \quad \Longleftrightarrow \quad \bm{u_0} = \sum_{i \in \text{stencil}} \bm{u_i} \delta(\bm{x_0} - \bm{x_i}) \Delta^3 ,
\end{equation}
where ``stencil" denotes the Eulerian cells which center $\bm{x_i}$ is such that $\delta(\bm{x_0} - \bm{x_i}) \neq 0$, and $\bm{u_i}$ is the velocity of a given fluid cell. Similarly, the membrane force $\bm{F_0}$ at a Lagrangian node is spread to a force density field $\bm{f}$ using:
\begin{equation}
  \label{eq:lag2eul}
  \mathcal{M}(\text{supp}(\delta)) \bm{f} = \int_\Omega \bm{F_0} \delta(\bm{x_0} - \bm{x}) d \bm{x} \Longleftrightarrow \bm{f_i} = \bm{F_0} \delta(\bm{x_0} - \bm{x_i}),
\end{equation}
where $\mathcal{M}(\text{supp}(\delta))$ is the measure of the support of the regularized Dirac-delta function, and is in practice equal to $64 \Delta^3$ in three dimensions for the regularized Dirac-delta function we choose in \refeq{eq:dirac-delta}.

Once the Lagrangian velocities of all the capsule nodes have been interpolated from the Eulerian velocity field using \refeq{eq:eul2lag}, the position of each node is updated using a second-order Runge-Kutta time integration scheme. The membrane stresses are then computed from the new configuration of the capsule using the methods described later in \refsec{sec:impl_mb_forces}, and transferred to the background fluid using \refeq{eq:lag2eul}. In the context of particle-laden flows, the use of the IBM requires sub-time stepping for the particle advection due to the orders of magnitude difference in the fluid time scale and the solid-solid interactions time scale \cite{Uhlmann2005}. This is not necessary for capsule-laden flows described by the FTM, i.e. the time step for the advection of Lagrangian nodes is equal to that of the fluid solver.
It follows that the trajectories of each Lagrangian node coincide with the streamlines of the flow and that the triangulations of two interacting capsules can never overlap $-$ provided that the time step is sufficiently small. As such, there is no need for any ad-hoc repulsive force between two approaching capsules or between a capsule approaching a wall, as was the case in e.g. \cite{lu2019scalable}: the non-penetration condition is seamlessly handled by the local flow field. In the latter case of close interaction between a capsule and a wall, however, the definition of the IBM stencil must be altered, as described in the next subsection.

Additionally, the case of capsules of inner viscosity $\mu_i$ different from the viscosity of the surrounding fluid $\mu_e$ needs special treatment. We adopt the approach developped in the original FTM by Unverdi \& Tryggvason \cite{unverdi1992front}: a discrete indicator function $I$ is computed from a discrete ``grid-gradient" field $\bm{G}(\bm{x})$:
\begin{equation}
  \bm{G}(\bm{x}) = \sum_{i \in \mathcal{T}} S_i \delta(\bm{x} - \bm{x}_i) \bm{n}_i,
\end{equation}
where $\mathcal{T}$ denotes the set of all triangles of the discretization of the surface of all the capsules, $S_i$ is the surface area of triangle $i$, $\bm{x}_i$ is the position vector of its centroid and $\bm{n}_i$ is its unit \textit{inward} normal vector. In practice, $\bm{G}(\bm{x})$ is computed by looping over all triangles of the discretizations of all capsules and spreading the quantity $S_i \bm{n}_i$ using the regularized Dirac-delta functions introduced previously. As such, $\bm{G}(\bm{x})$ is non-zero in the union of all the IBM stencils. The discrete indicator function $I(\bm{x})$ is computed by solving the following Poisson problem:

\begin{equation}
  \Delta I = \nabla \cdot \bm{G}.
\end{equation}

Since $I$ is a regularized step function, it should have constant values away from the capsule membranes. To guarantee this property, we only update $I$ in the cells where $\bm{G}$ is non-zero and we re-initialize $I$ to $0$ or $1$ elsewhere, as suggested in \cite{tryggvason2001front}.

\subsubsection{Adaptive FTM strategy \label{sec:FTM-adapt}}
Our current implementation of the FTM requires all cells in an IBM stencil to be the same size $\Delta$. As a results, all the Eulerian cells around the membrane must be the same size as well, although future studies may lift this restriction. In practice, since the flow physics is happening in the vicinity of the capsule, we need the cell sizes around the membrane to be the smallest grid size of the fluid domain. To enforce this condition, we create a scalar field $\xi$ initialized at each time step to be: (i) $0$ if the Eulerian cell does not belong to any IBM stencil; or (ii) a randomly generated value between $0$ and $1$ otherwise. In other words, the scalar field $\xi$ tags the IBM stencils with noise while the rest of the domain is set to a constant value. Feeding this scalar field to Basilisk's wavelet adaptation algorithm ensures that all the stencil cells are defined at the finest level, and that no IBM stencil contains Eulerian cells of different cell levels.

\subsection{Computation of the membrane forces\label{sec:impl_mb_forces}}
\subsubsection{Computation of the elastic force with the Finite Element Method}
\label{sec:fem}
In order to compute the nodal elastic forces given by \refeq{eq:f_el}, we employ a Finite Element Method (FEM). In most FEM solvers from Engineering applications, the sought quantity is the displacement of a structure under a known applied stress. In the case of biological membranes, we rather seek the internal stress of the membrane under a known displacement \cite{barthes2016motion}. Charrier et al. \cite{charrier1989free} have been the first to design this specific FEM framework: we base our implementation on their work as well as that of Doddi \& Bagchi \cite{doddi2008lateral}.

Consider an arbitrary triangle $T_i$ on the discretized membrane: in order to compute the elastic force of its three vertices, we first rotate it to a common plane $-$ e.g. the $x,y$-plane $-$ using a rotation matrix $\bm{R_i}$ from the current orientation of the triangle to its orientation in the common plane. Then, we assume the position of the triangle vertices in a stress-free configuration is known in the common plane, and we compute the displacements $\bm{v_k}$ of each of the three vertices of $T_i$. Using linear shape functions, the deformation gradient tensor and the Cauchy-Green deformation gradient tensor attached to $T_i$ can be computed:
\begin{equation}
  \bm{F} = \frac{\partial \bm{v_k}}{\partial \bm{x^p}}, \qquad \bm{C} = \bm{F^T}\bm{F},
  \label{eq:fc}
\end{equation}
where $(\bm{x^p}, \bm{y^p})$ is the basis of the common plane. Note that $\bm{F}$ and $\bm{C}$ are two-dimensional tensors and correspond to the tangential components of $\bm{F_s}$ and $\bm{C_s}$ in \refeqs{eq:fs}{eq:cs}. By diagonalizing $\bm{C}$ and taking the square root of its eigenvalues ($\bm{C}$ is symmetric positive definite), we can access the two principal stretch ratios $\lambda_1$, $\lambda_2$ attached to $T_i$. Following Charrier et al. \cite{charrier1989free}, the principle of virtual work yields the expression linking the nodal force and nodal displacement at node $j$:
\begin{equation}
  \bm{F^P_{\text{elastic},j}} = A_i \frac{\partial W}{\partial \lambda_1} \frac{\partial \lambda_1}{\partial \bm{v_j}} + A_i \frac{\partial W}{\partial \lambda_2} \frac{\partial \lambda_2}{\partial \bm{v_j}},
  \label{eq:ref-nodal-force}
\end{equation}
where $A_i$ is the area of $T_i$. Rotating \refeq{eq:ref-nodal-force} back to the current reference frame of $T_i$, we get the final expression of the contribution of triangle $T_i$ to the elastic force of node $j$:
\begin{equation}
  \begin{split}
    \bm{F_{\text{elastic},j}} & = \bm{R^T} \bm{F^P_{\text{elastic},j}}
     \\ & = A_i \bm{R^T} \left( \frac{\partial W}{\partial \lambda_1} \frac{\partial \lambda_1}{\partial \bm{v_j}} + \frac{\partial W}{\partial \lambda_2} \frac{\partial \lambda_2}{\partial \bm{v_j}} \right).
     \label{eq:nodal-force}
   \end{split}
\end{equation}
This FEM implementation is summarized in algorithm \ref{alg:fem}.

\begin{algorithm}
\caption{Pseudocode for the Finite Element Method}\label{alg:fem}
\begin{algorithmic}
  \Loop{ over all triangles $i$}
      \Loop{ over the three nodes $j$ of $T_i$}

          Compute $\bm{x^P_j} = \bm{R} \bm{x_j}$

          Compute the nodal displacement $\bm{v_j} = \bm{x^P_j} - \bm{x^P_{j,\,t=0}}$
      \EndLoop

      Compute $\bm{F}, \, \bm{C}$ from \refeq{eq:fc}

      Compute the eigenvalues of $\bm{C}$ and $\bm{F}$, i.e. $\lambda_1^2, \, \lambda_2^2, \, \lambda_1, \, \lambda_2$

      \Loop{ over the three nodes $j$ of $T_i$}

          Compute $\partial \lambda_1/\partial \bm{v_j}$, $\partial \lambda_2/\partial \bm{v_j}$

          Compute $\bm{F_{\text{elastic},j}^P}$ from \refeq{eq:ref-nodal-force}

          Rotate $\bm{F_{\text{elastic},j}^P}$ to the current orientation of $T_i$

          Add $\bm{F_{\text{elastic},j}}$ to the total elastic force of node $j$
      \EndLoop
  \EndLoop
\end{algorithmic}
\end{algorithm}

\subsubsection{Computation of the bending force using paraboloid fits}
\label{sec:bending-implementation}
The computation of $\bm{F}_{\text{bending}}$ relies on the local evaluation of: (i) the mean and Gaussian curvatures $\kappa$ and $\kappa_g$, (ii) the Laplace-Beltrami operator of the mean curvature $\Delta_s \kappa$, and (iii) a relevant control area $A$.

To evaluate $\kappa$ and $\kappa_g$ at node $i$, we blend the approaches of Farutin et al. \cite{farutin20143d} and Yazdani \& Bagchi \cite{yazdani2012three}. A local reference frame is attached to node $i$, with the $z$-direction coinciding with the approximate normal vector $\bm{n_i}$. Then, a paraboloid is fitted to node $i$ and its one-ring neighbors. In our triangulated surface, most nodes have six neighbors\footnote{Exactly twelve nodes have five neighbors, since we discretize a spherical membrane by subdividing each triangle of an icosahedron. Each newly created node is projected back to a sphere, and if necessary projected onto a more complex shape, e.g. a biconcave membrane.}, making the system overdetermined and a least-squares method is used. From this paraboloid fitting, we can derive the local mean and Gaussian curvatures $-$ see equations (12) and (13) in \cite{yazdani2012three} $-$, as well as a refined approximation of $\bm{n_i}$. This procedure is iterated using the newest normal vector approximation to define the local frame of reference, until satisfactory convergence of $\bm{n_i}$ is reached. Our numerical experimentations show that between three to five iterations usually suffice to obtain a converged normal vector.

The same paraboloid fitting method is used to compute $\Delta_s \kappa$, or $\Delta_s (\kappa - \kappa_0)$ in the case of non-zero reference curvature. This time, a paraboloid is fitted to the \textit{curvatures} of node $i$ and its neighbors, and then differentiated to obtain the desired surface Laplacian.

The last term $A$ is necessary to obtain a bending force as opposed to a bending force per surface area. Let $A_i$ denote the nodal area attached to node $i$: at any time the sum of all nodal areas need to equal the total area of the capsule, i.e. $\sum_{i=0}^N A_i = A_{tot}$ with $N$ the number of Lagrangian nodes and $A_{tot}$ the total area of the discretized surface of the capsule. The Voronoi area of node $i$ enforces this property only for non-obtuse triangles. As such, we adopt the ``mixed-area" of Meyer et al. \cite{guckenberger2016bending, meyer2003discrete} which treats the special case of obtuse triangles separately: if a triangle $j$ is not obtuse, its contribution to the nodal area of its vertices is the standard Voronoi area; while if $j$ is an obtuse triangle, the nodal area of its obtuse vertex is $A_j/2$ while the nodal area of the remaining two vertices is $A_j/4$, where $A_j$ is the area of triangle $j$.

\section{Validation Cases \label{sec:validation}}
\subsection{Elastic and bending forces of an isolated membrane}
\subsubsection{Elongation of a flat elastic membrane}
Our first validation case focuses on the computation of the elastic stress in the membrane. To this end, we stretch an isolated flat membrane devoid of bending resistence in one of its principal direction $\bm{e_1}$ while ensuring the principal stress $T_2$ in the second principal direction $\bm{e_2}$ remains zero. We then analyze the non-zero principal stress $T_1 = (\partial W / \partial \lambda_1)/\lambda_2$ as a function of the principal stretch $e_1 = (\lambda_1^2 - 1)/2$.
This test is repeated for two membranes: the former obeying the neo-Hookean law and the latter obeying the Skalak law. Note that in order to set the principal stress $T_2$ equal to zero, we impose $\lambda_2$ to a value strictly lower than $1$, i.e. the membrane is shrinked in the second principal direction:

\begin{equation}
  \lambda_2 = \begin{cases}
    1/\sqrt{\lambda_1} \qquad \text{for the neo-Hookean law}\\
    \sqrt{(1 + C\lambda_1^2)/(1 + C\lambda_1^4)} \qquad \text{for the Skalak law}
  \end{cases}
\end{equation}

\begin{figure}
  \centering
  \includegraphics[height=.4\textheight]{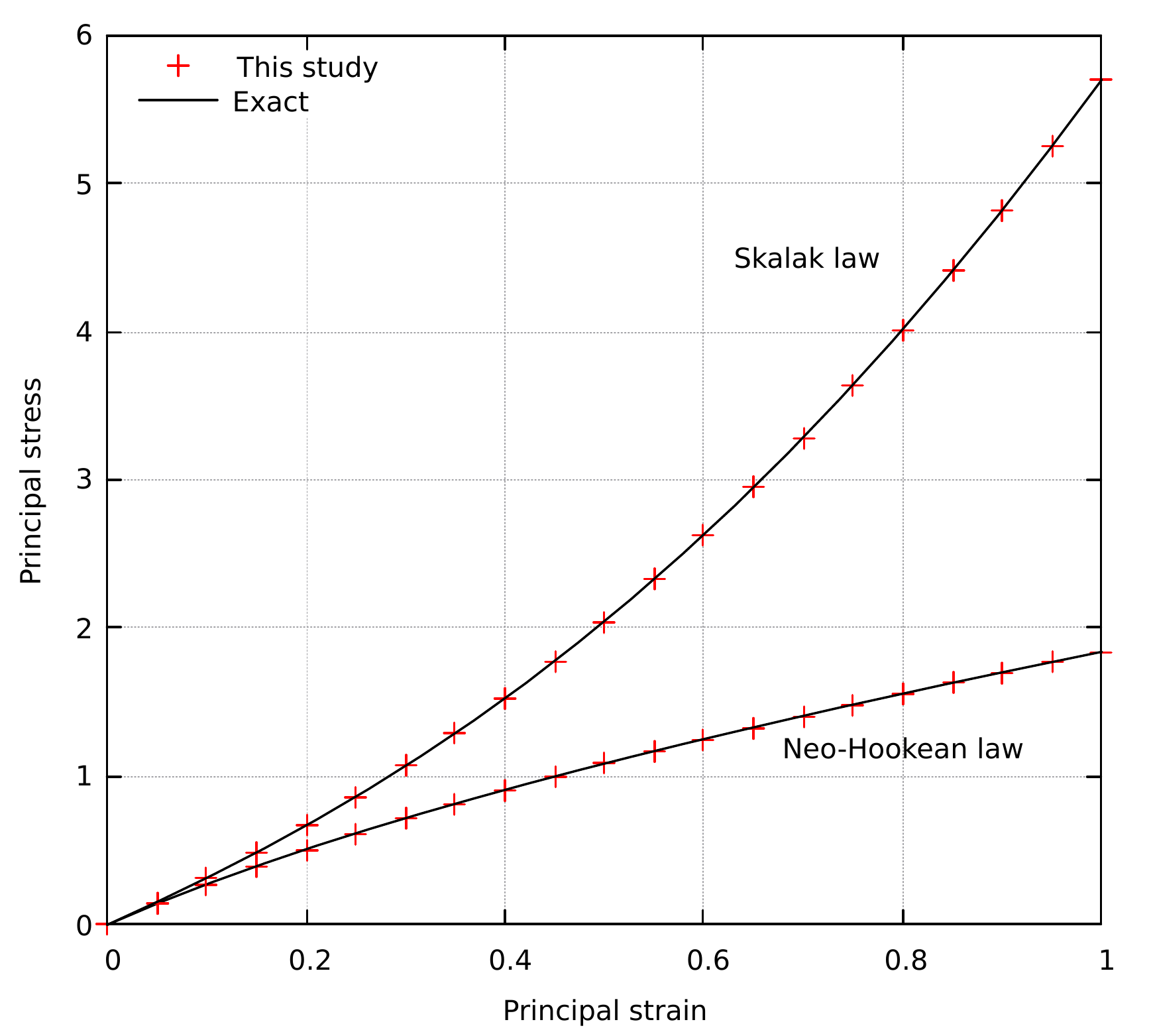}
  \caption{Stress-strain response of an isolated flat membrane for the neo-Hookean and Skalak elastic laws. The results from this study are compared to exact expressions derived in \cite{barthes2002effect}.}
  \label{fig:flat-mb-elongation}
\end{figure}

We compare our results to the exact stress derived by Barthès-Biesel et al. \cite{barthes2002effect} in \reffig{fig:flat-mb-elongation}, with $E_s = C = 1$. The data we generate overlaps perfectly with the analytical stress-strain relations, thus validating the implementation of our Finite-Element solver for the elastic membrane stresses. The source code to reproduce this validation case is available online \cite{uniaxial_stretch.c}.

\subsubsection{Bending force of a curved membrane}
\begin{figure}
  \centering
  \includegraphics[width=.65\columnwidth]{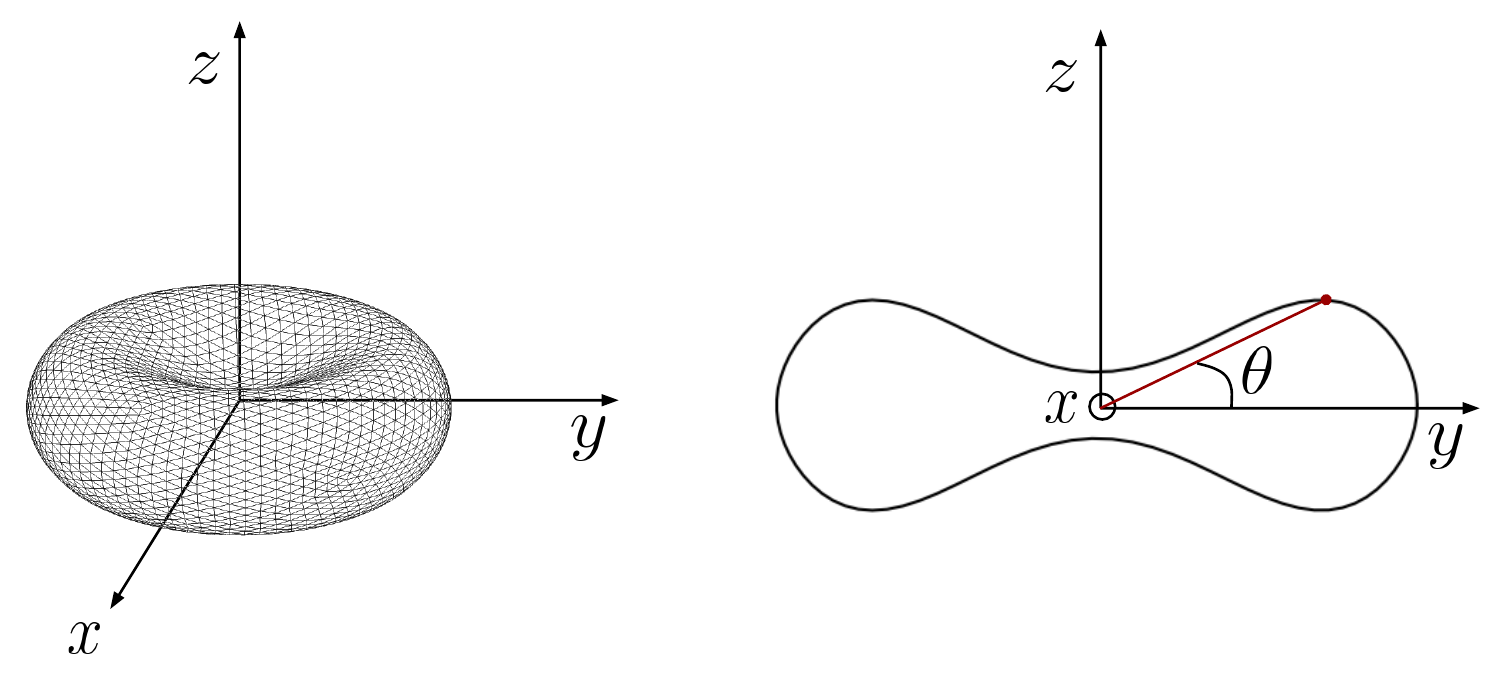}
  \caption{Schematic of a biconcave capsule centered at the origin, and definition of the polar angle $\theta$.}
  \label{fig:biconcave-angle}
\end{figure}

In order to validate our bending force, we follow the procedure of Guckenberger et al. \cite{guckenberger2016bending}: considering a biconcave membrane with zero reference curvature, we compare the mean and Gaussian curvatures, Laplace-Beltrami operator of the mean curvature, and total nodal bending force density to analytical expressions derived using a symbolic calculus software. Since the biconcave capsule has a rotational symmetry around the $z$-axis and a symmetry with respect to the $(x,y)$-plane, we plot our results according to the angle $\theta$ defined in \reffig{fig:biconcave-angle}, with $\theta$ varying from $0$ to $\pi/2$. This biconcave shape is a good candidate to test the bending force since its two principal curvatures are in general not equal to each other and are varying along the surface of the biconcave shape $-$ even changing sign. The following results are obtained with a biconcave membrane discretized by a triangulation containing $5120$ triangular elements.

\begin{figure}
  \centering
  \begin{subfigure}{.45\columnwidth}
    \hspace{-3cm}\includegraphics[height=.42\textheight]{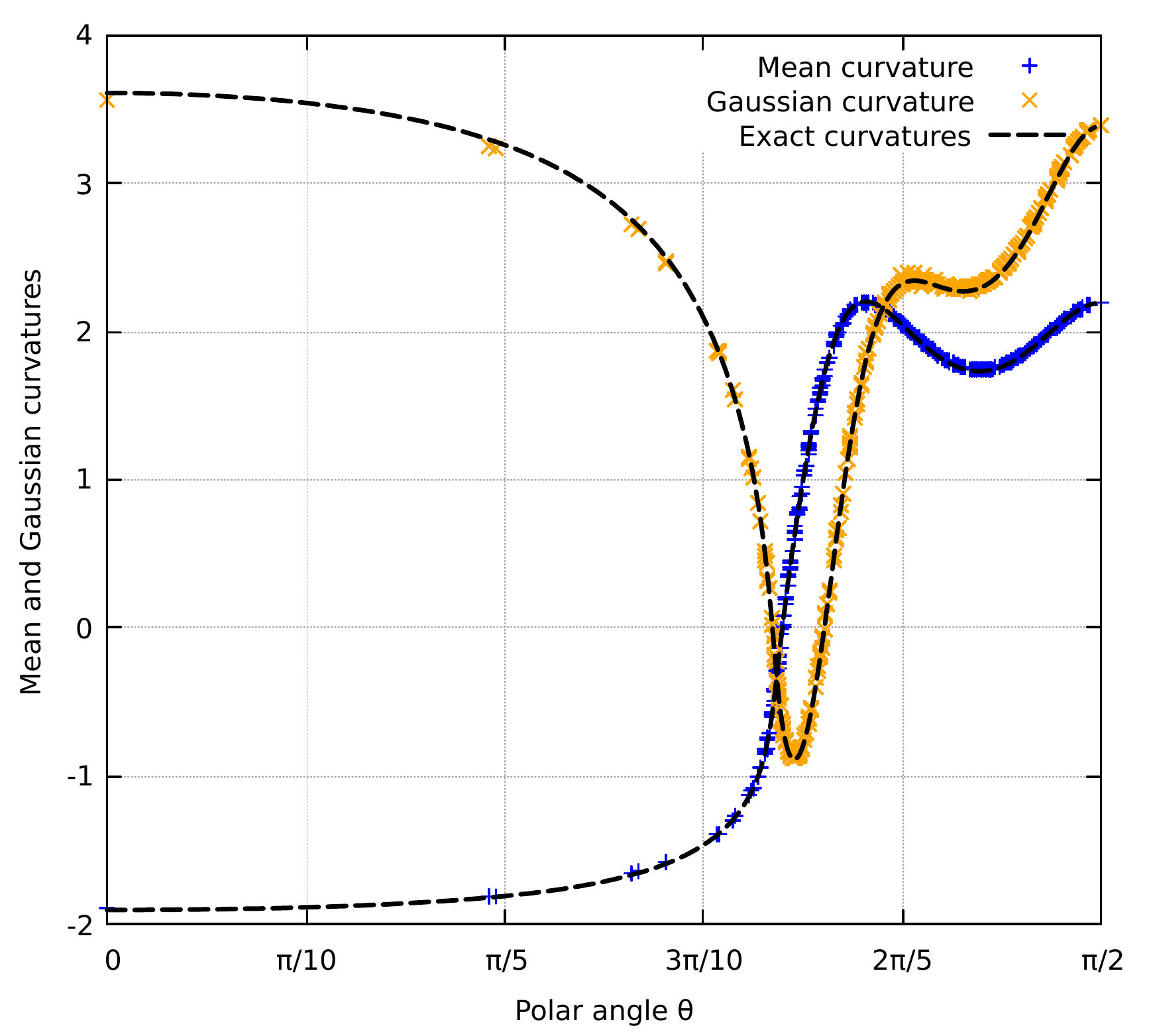}
    \subcaption{}
    \label{fig:curvature-validation-a}
  \end{subfigure}
  \hfill
  \begin{subfigure}{.45\columnwidth}
    \centering
    \includegraphics[height=.42\textheight]{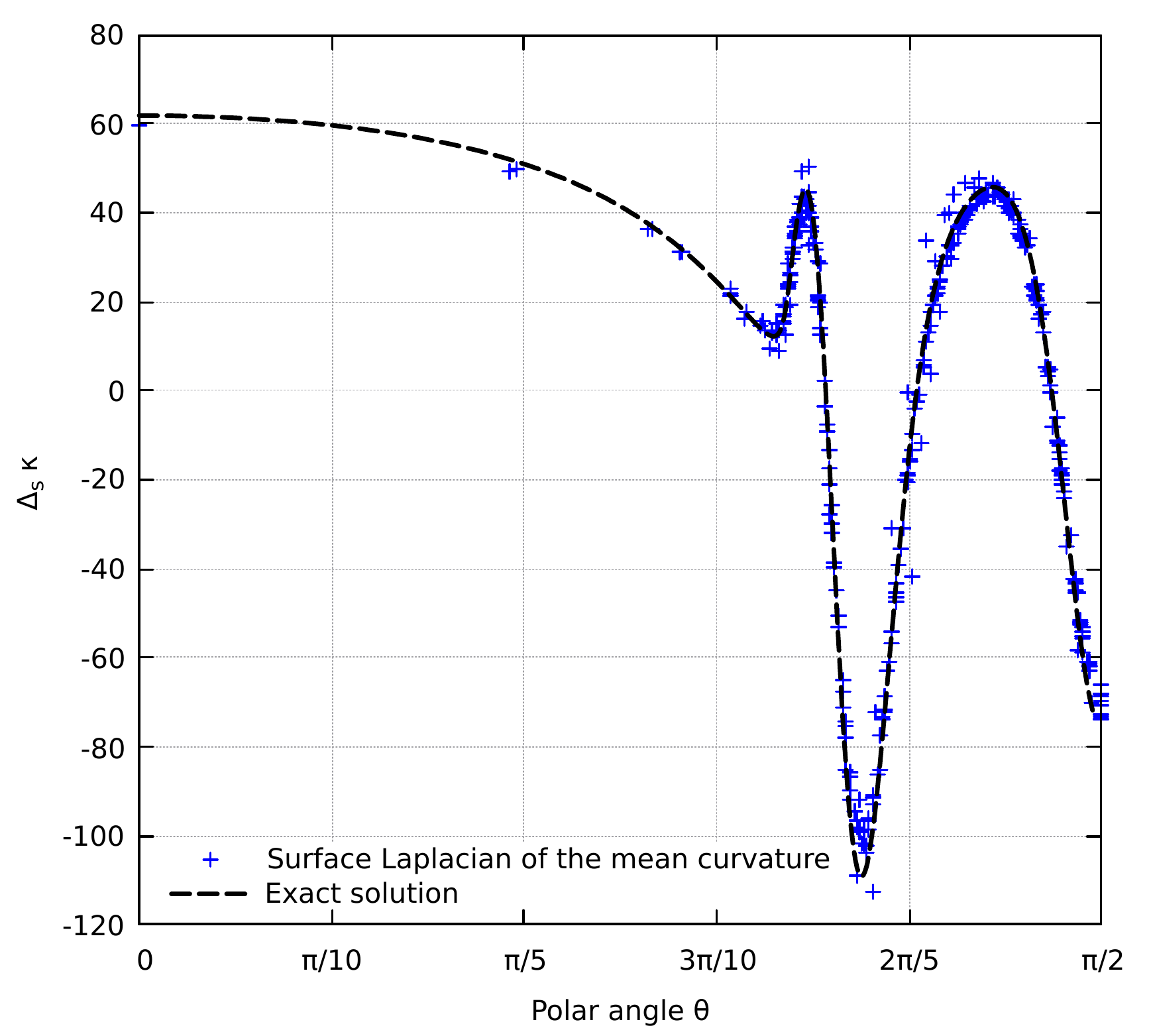}
    \subcaption{}
    \label{fig:curvature-validation-b}
  \end{subfigure}
  \begin{subfigure}{\columnwidth}
    \centering
    \includegraphics[height=.42\textheight]{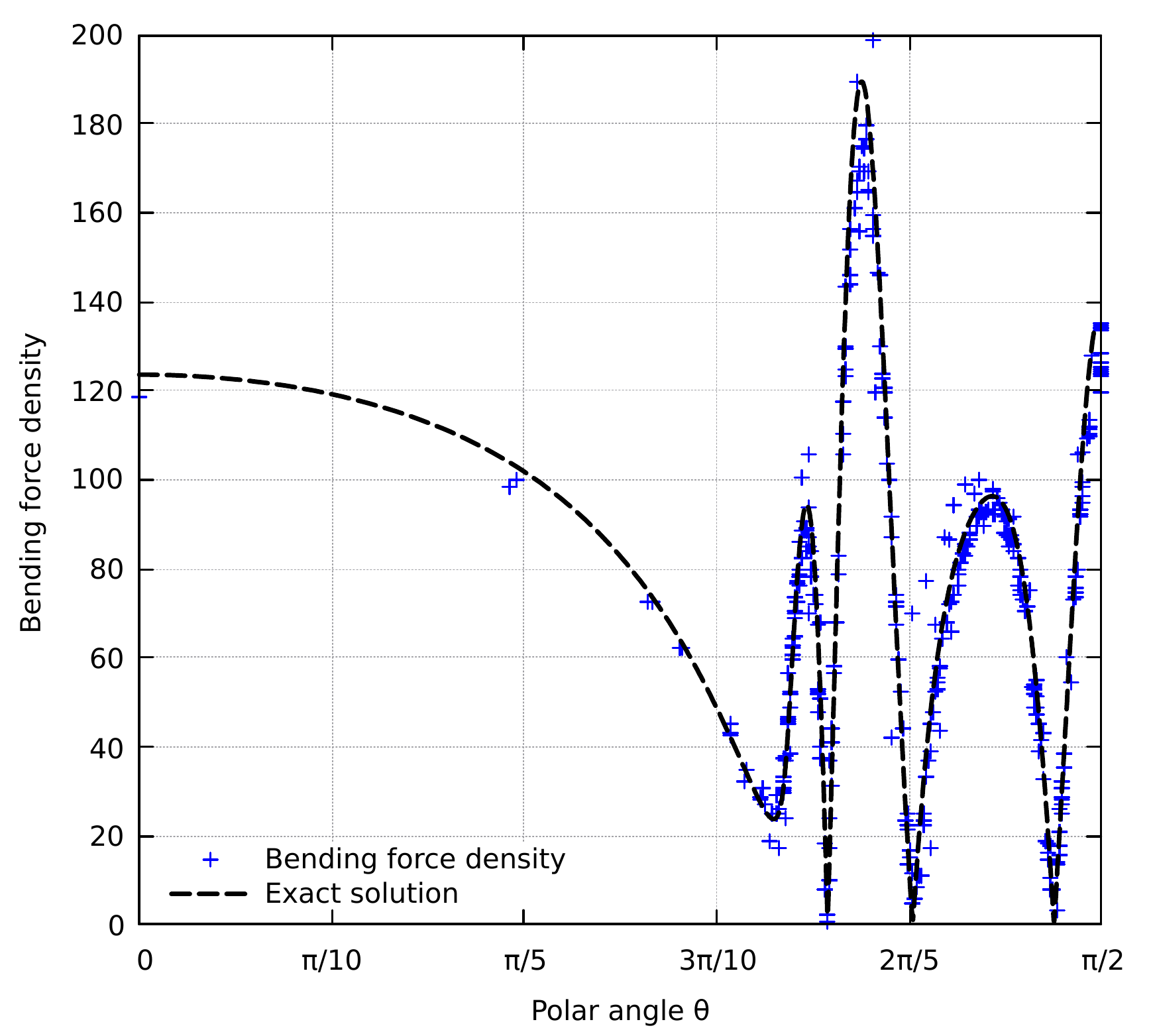}
    \subcaption{}
    \label{fig:curvature-validation-c}
  \end{subfigure}
  \caption{Comparison of the computed mean and Gaussian curvatures (top left), Laplace-Beltrami operator of the mean curvature (top right) and nodal bending force density (bottom) to their analytical expressions. All quantities are plotted against the polar angle $\theta$ defined in \reffig{fig:biconcave-angle}.}
  \label{fig:curvature-validation}
\end{figure}

We compare our computed mean and Gaussian curvatures to their respective analytical expressions in \reffig{fig:curvature-validation-a}: the agreement is very satisfactory for both curvatures. \Reffig{fig:curvature-validation-b} shows the Laplace-Beltrami operator of the curvature against its analytical expression: the general trend still matches that of the analytical expression very well, but a few outliers deviate from it by a few percents. The same behavior is observed in \reffig{fig:curvature-validation-c} which shows the nodal bending force density. The fact that the behavior of \reffig{fig:curvature-validation-b} and \reffig{fig:curvature-validation-c} is similar is note surprising, as the nodal bending force density plotted in \reffig{fig:curvature-validation-c} directly involves the Laplace-Beltrami operator of the mean curvature shown in \reffig{fig:curvature-validation-b}. It is expected to see some small deviations to the theory when taking the Laplace-Beltrami operator of the mean curvature, as we are essentially taking a fourth-order derivative of the geometry of the membrane, and Guckenberger et al. \cite{guckenberger2016bending} observe a similar noise when performing the same tests (see figures 6c and 8e in \cite{guckenberger2016bending}). In fact, they show that most other methods perform much worse at computing the Laplace-Beltrami operator of the mean curvature, and hence at computing the total bending force. As such, our implementation of the bending force shows the expected performance. The code to reproduce this test case is available at \cite{biconcave_curvatures.c}.

\subsection{Initially spherical capsule in an unbounded shear flow}
\subsubsection{Neo-Hookean elasticity without bending resistance}

\begin{figure}
  \centering
  \includegraphics[height=.47\textheight]{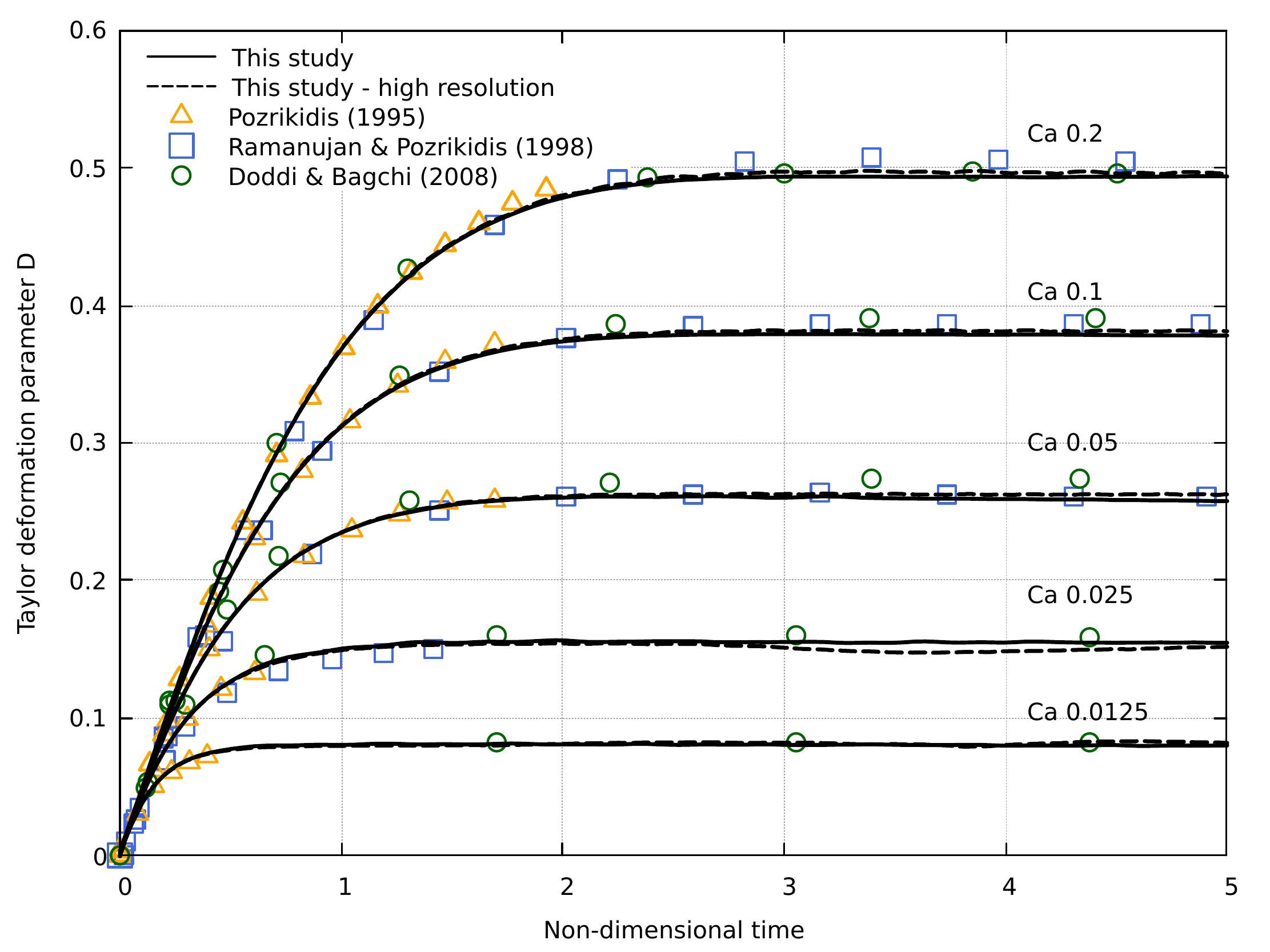}
  \caption{Taylor deformation parameter as a function of the non-dimensional time of an initially spherical Neo-Hookean capsule in an unbounded shear flow. The solid line corresponds to 32 Eulerian cells per initial diameter and a Lagrangian discretization using 1280 triangles, while the dashed line corresponds to 64 Eulerian cells per initial diameter and a Lagrangian discretization using 5120 triangles.}
  \label{fig:nh_shear_taylor}
\end{figure}

We now seek validation of the coupling between the membrane solver and the fluid solver. To this end, we consider an initially spherical capsule of radius $a$ in an unbounded shear flow. The elasticity is governed by the neo-Hookean law, and the flow field is initialized to be that of an undisturbed shear flow. As the capsule deforms, we plot the Taylor deformation parameter $D = (a_{max} - a_{min})/(a_{max} + a_{min})$ as a function of the non-dimensional time $\dot{\gamma} t$, with $a_{max}$ and $a_{min}$ the maximum and minimum radii of the capsule at a given time, and $\dot{\gamma}$ the shear rate. We perform this simulation for various Capillary numbers $Ca = \mu a^2 \dot{\gamma}/E_s$, with $E_s$ the elastic modulus. In this test case, Stokes conditions are intended so we set the Reynolds number to 0.01. At time $t = 0$, the flow field is set to that of a fully developped shear flow: $u_x = \dot{\gamma} y$. The computational box is bi-periodic in the x and z directions, while Dirichlet boundary conditions for the velocity are imposed in the y direction. The length of the computational box is equal to 8 initial radii, the size of the most refined Eulerian cells is set to $1/128$ that of the domain length, and the membrane is discretize by 1280 triangles.
The non-dimensional time step $\dot{\gamma}\Delta t$ is set to $10^{-3}$, except for the Capillary numbers $Ca = 0.025$ and $Ca = 0.0125$ where the time step is decreased to $\dot{\gamma}\Delta t = 10^{-4}$ to stabilize the elastic force computation.

\begin{figure}
  \centering
  \includegraphics[height=.3\textheight,trim=150 150 150 150, clip]{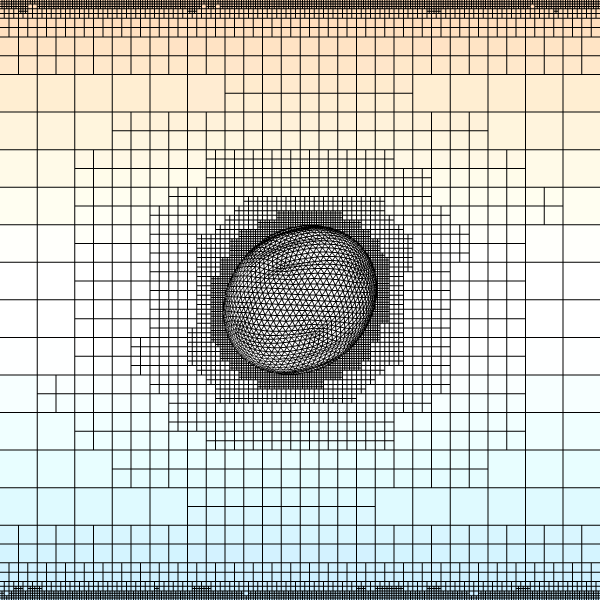}
  \caption{A zoomed in snapshot of a buckling membrane at $Ca = 0.0125$ with a Lagrangian discretization comprising 5120 triangles. This behavior is arising due to the absence of bending stresses, and the buckling wavelength is dependant on the Lagrangian discretization. The color field represents the $x$-component of the velocity.}
  \label{fig:buckling}
\end{figure}

This case has been widely studied in the literature: in \reffig{fig:nh_shear_taylor} we compare our results to those of \cite{pozrikidis1995finite, ramanujan1998deformation} who used the BIM, as well as \cite{doddi2008lateral} who used the FTM. The agreement is very satisfactory: the steady-state value we obtain for $D$ is well within the range of the reported data, both in the transient regime and once a steady-state is reached.
We also show in \reffig{fig:nh_shear_taylor} the results for a finer triangulation of the membrane with 5120 triangles, and a refined Eulerian mesh with the finest Eulerian cell size corresponding to $1/256$ that of the domain length. The only difference is that the steady state is longer to reach for $Ca = 0.025$ and $Ca = 0.0125$, due to the apparition of buckling instabilities on the membrane as a result of the absence of bending stresses. This buckling instability has been observed both experimentally \cite{koleva2012deformation} and numerically \cite{walter2010coupling, yazdani2013influence}, although in numerical simulations the wavelength is unphysical and determined by the size of the mesh discretizing the capsule \cite{barthes2016motion}. In our simulations, we do observe the same dependance of the wavelength of the membrane buckles on the Lagrangian mesh element size.  An example of this buckling instability is shown in \reffig{fig:buckling}. The code to reproduce this test case is available at \cite{nh_shear.c}.

\subsubsection{Including bending resistance}
\begin{figure}
  \centering
  \includegraphics[height=.5\textheight]{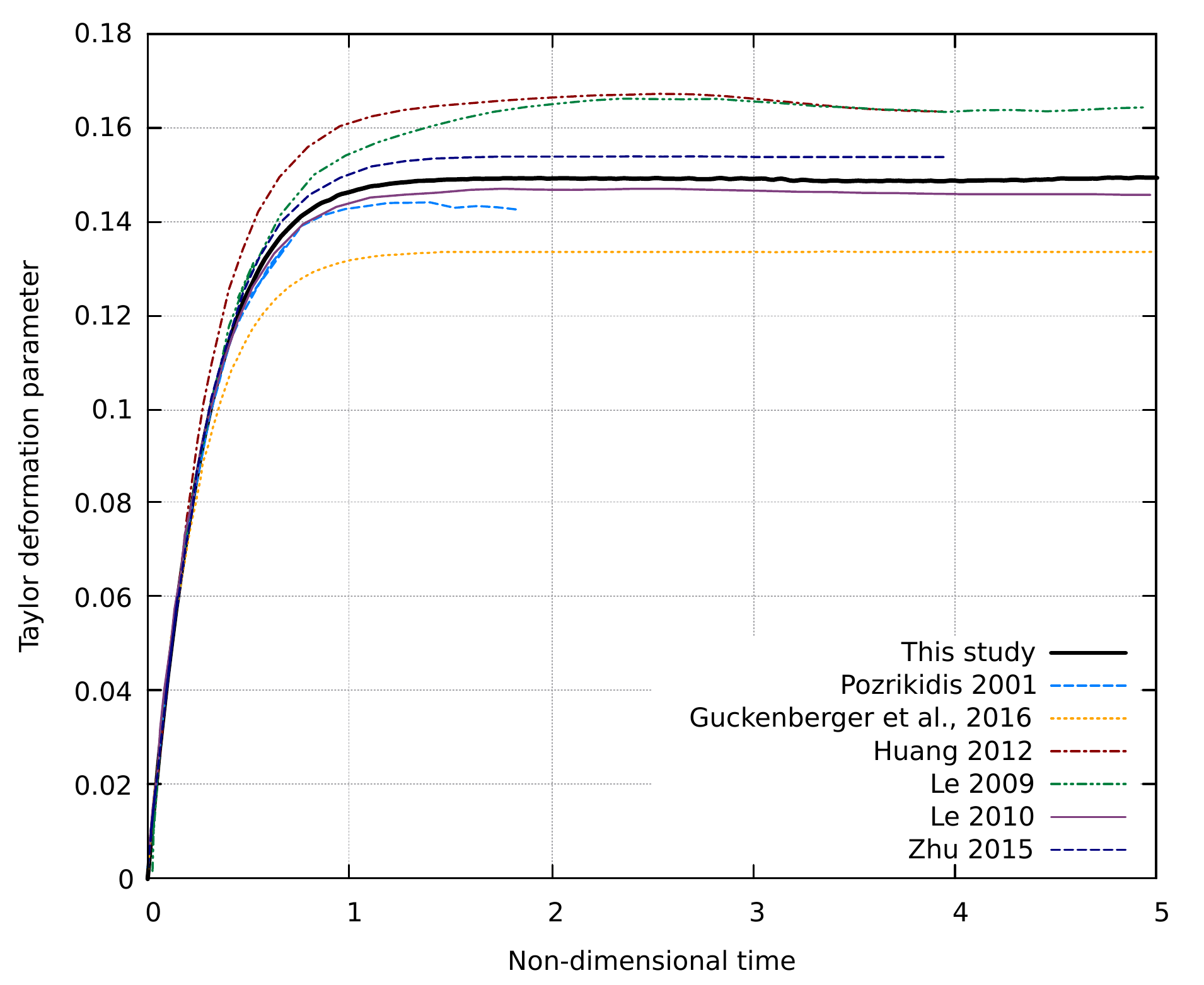}
  \caption{Taylor deformation parameter of an isolated capsule undergoing elastic and bending stresses in a shear flow. The capillary number is $Ca = 0.05$ and the non-dimensional bending coefficient is $\tilde{E_b} = 0.0375$.}
  \label{fig:bending_shear}
\end{figure}

We further validate our solver by considering the similar case of a capsule deforming in a shear flow, this time with the addition of a bending force. As in the previous case, an initially spherical, unstressed capsule is placed in a shear flow where the initial velocity field is fully developped. The Capillary number is $Ca = 0.05$, and the non-dimensional bending coefficient $\tilde{E_b} = E_s/(a^2 E_b)$ is chosen equal to $0.0375$. The membrane is discretized with 5120 triangles and the same Eulerian resolution as in the previous case is chosen. Due to the stiffness of the bending force, we set the time step to $\Delta t = 10^{-4}$. The Taylor deformation parameter is compared to that of various studies in the literature in \reffig{fig:bending_shear}. The capsule deforms under the action of the flow field and the Taylor deformation parameter quickly attains a steady state of about $D = 0.15$. We remark that the data reported in the literature is scattered by about 20\% which underlines the challenges to simulate Helfrich's bending force, as was previously noted by \cite{guckenberger2016bending}.
We also note that our results are situated well within the range of the reported data: we are close to the results of Zhu \& Brandt \cite{zhu2015motion} and Le et al. \cite{le2010effect}, and our curve is located in the middle of the reported range that we borrowed from \cite{guckenberger2016bending, guckenberger2017theory}. Given such a wide range of reported literature data, it is difficult to conduct a rigorous quantitative analysis. Nevertheless, we conclude from \reffig{fig:bending_shear} that our bending force shows a similar behavior as that of other studies, a claim also supported by the validation case in \refsec{sec:rbc_shear}. The code to reproduce this test case is available at \cite{bending_shear.c}

\subsection{Initially spherical capsule flowing through a constricted channel}
\label{sec:constricted_channel}
To validate our implementation for an elastic capsule in the presence of complex boundaries, we consider the case of a capsule flowing through a constricted square channel proposed by Park \& Dimitrakopoulos \cite{park2013transient}. The elasticity of the membrane is governed by the Skalak law with the area dilatation modulus $C$ set to 1, the capsule is initially pre-inflated such that its circumference is increased by $5\%$, and the flow is driven by an imposed uniform velocity field at the inlet and outlet boundaries. We follow \cite{park2013transient} and choose the Capillary number to be $0.1$, and since Stokes conditions are intended we set $Re = 0.01$.

\begin{figure}
    \centering
    \includegraphics[height=.8\textheight]{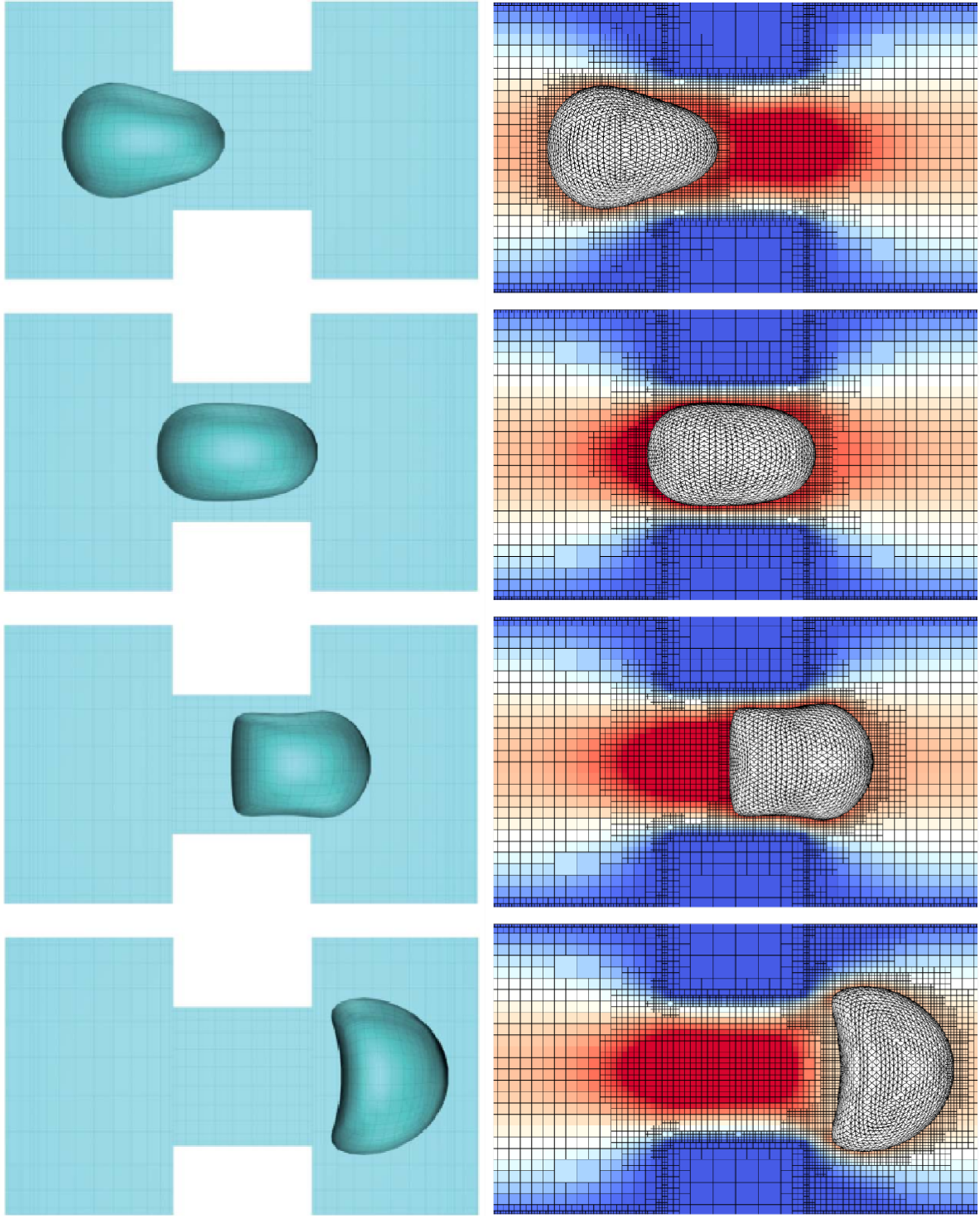}
  \caption{Snapshots of the capsule as it flows through the constriction. Left: Park \& Dimitrakopoulos \cite{park2013transient}; Right: this study.}
  \label{fig:const_channel_snap}
\end{figure}

\begin{figure}
  \centering
  \includegraphics[height=.4\textheight, trim=0 0 0 40, clip]{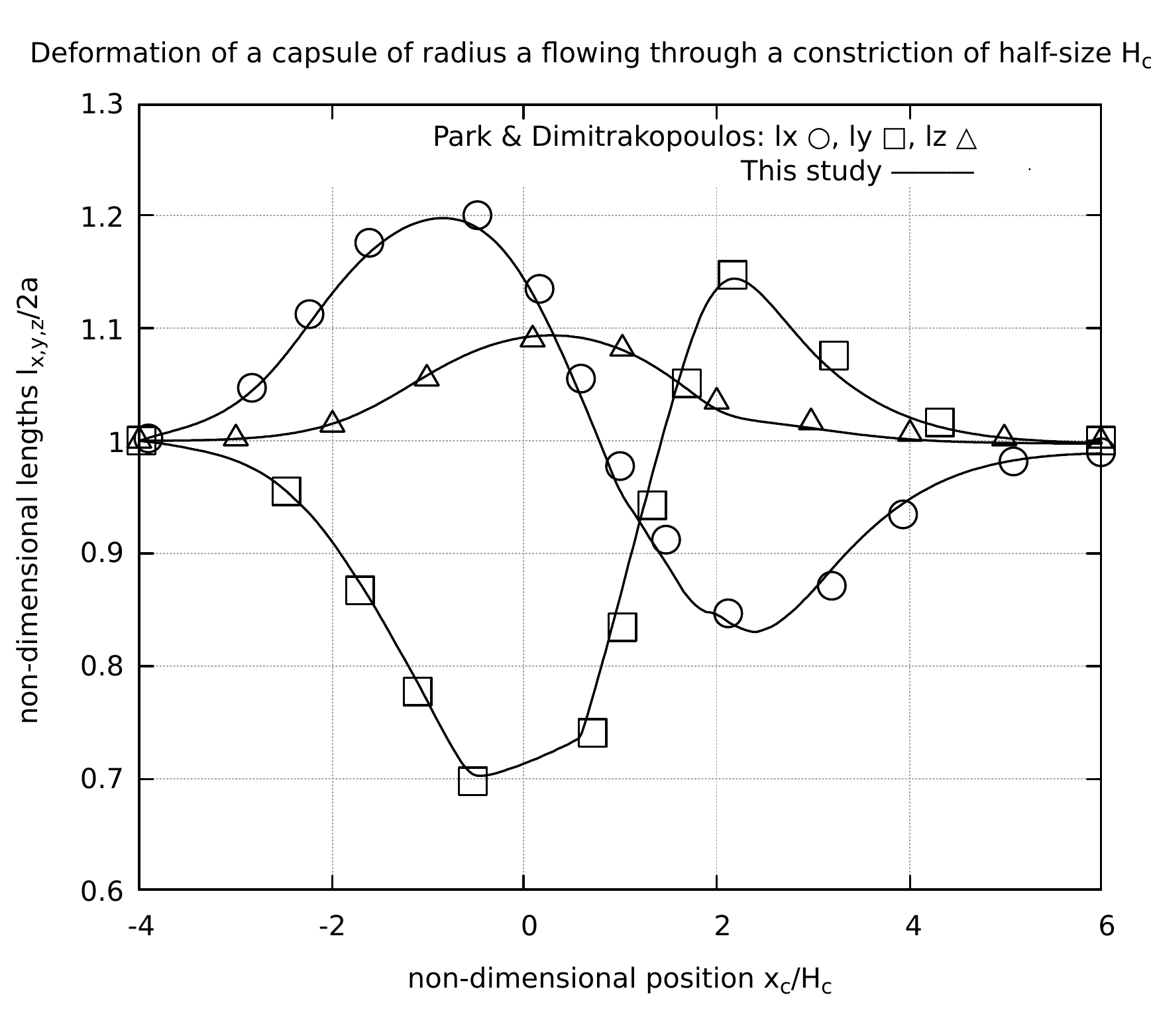}
  \caption{Non-dimensional lengths of the capsules in the three directions $x$, $y$ and $z$ with respect to the non-dimensional $x$-position of the center of the capsule. Results are compared to \cite{park2013transient}.}
  \label{fig:const_channel_lxyz}
\end{figure}

The results are presented in \reffig{fig:const_channel_snap} and \reffig{fig:const_channel_lxyz}. The qualitative agreement in \reffig{fig:const_channel_snap} is very satisfactory as the capsule shape is visually identical to that of \cite{park2013transient}. We draw the reader's attention to the adaptive Eulerian mesh on the right-hand side of \reffig{fig:const_channel_snap}: the cells size is imposed to be minimal at the solid boundaries and around the capsule, while everywhere else the adaptivity criterion is governed by the velocity gradients. As a result, the grid cells away from the membrane and from the walls quickly coarsen to up to three levels lower, except in the vicinity of the corners where stronger velocity gradients occur. \Reffig{fig:const_channel_lxyz} shows the non-dimensional lengths of the capsule in the $x$-, $y$- and $z$-directions with respect to the non-dimensional position of its center $x_c/H_c$, with $x_c$ and $H_c$ the $x$-position of the center of the capsule and the half-height of the constriction, respectively. As found by Park \& Dimitrakopoulos, the final shape of the capsule is not exactly spherical as it remains shrinked in the $x$-direction downstream of the constriction. Despite some small deviations during the extreme deformation of the capsule, around $x_c/H_c = -1$, the overall quantitative agreement of the transient shape of the capsule is also satisfactory, especially considering that other authors have reproduced this case with similar or larger deviations from the results reported by Park \& Dimitrakopoulos \cite{balogh2017computational, ii2018continuum}. The code to reproduce this test case is available at \cite{constricted_channel.c}.

\subsection{Red blood cell in an unbounded shear flow \label{sec:rbc_shear}}
The next test case aims at validating the membrane solver when a viscosity ratio $\lambda_\mu = \mu_i / \mu_e$ is different than $1$. To this end, we consider an RBC in an unbounded shear flow, with $\lambda_\mu = 5$. The membrane forces include the Skalak elastic law and the Helfrich's bending force. The Capillary number is $Ca = 0.1$, the area dilatation modulus $C$ is chosen equal to $50$ and the non-dimensional bending coefficient is $\tilde{E_b} = 0.01$.
The reference curvature is $c_0 a = -2.09$ \cite{pozrikidis2005resting, yazdani2011phase}, where $a = (3V/4\pi)^{1/3}$ is the radius of the sphere of equal volume as that of the RBC. The initial shape of the RBC is biconcave and is described by the following equations, for an RBC which largest radius is orthogonal to the $y$ direction \cite{pozrikidis2010computational}:

\begin{equation}
  \label{eq:rbc_shape}
  \begin{cases}
    x = a c \cos\phi \sin\psi\\
    y = \frac{a c}{2} \sin\phi \left( \alpha_1 + \alpha_2 \cos^2\phi + \alpha_3 \cos^4\phi \right)\qquad\qquad \text{with} \; \phi \in [0, 2\pi], \; \psi \in [-\frac{\pi}{2}, \frac{\pi}{2})\\
    z = a c \cos\phi \sin\psi,
  \end{cases}
\end{equation}

with $\alpha_1 = 0.207$, $\alpha_2 = 2.003$ and $\alpha_3 = -1.123$. Since we consider a viscosity ratio, we also define the initial indicator function $I$ as the volume fraction of inner fluid:

\begin{equation}
  I(\bm{x}) = \begin{cases}
    1 \; \text{if} \; \Phi(\bm{x}) < 0\\
    0 \; \text{if} \; \Phi(\bm{x}) > 0\\
    \text{between 0 and 1 otherwise},
  \end{cases}
\end{equation}

where $\Phi$ is the level-set alternative formulation of \refeq{eq:rbc_shape}:

\begin{equation}
  \Phi(x, y, z) = \frac{x^2 + z^2}{(a c)^2} + \frac{4 y^2}{(a c)^2} \left( \alpha_1 + \alpha_2\frac{x^2 + z^2}{(a c)^2} + \alpha_3\left(\frac{x^2 + z^2}{(a c)^2}\right)^2 \right).
\end{equation}

The initial fluid velocity is set to that of an unbounded shear flow of shear rate $\dot{\gamma}$ with the velocity gradient in the direction of the greater axis of the RBC. The dimensionless time step we use is $\dot{\gamma}\Delta t = 10^{-4}$ and is determined from trial and error.

\begin{figure}
  \centering
  \includegraphics[height=.16\columnwidth, trim=130 130 130 130, clip]{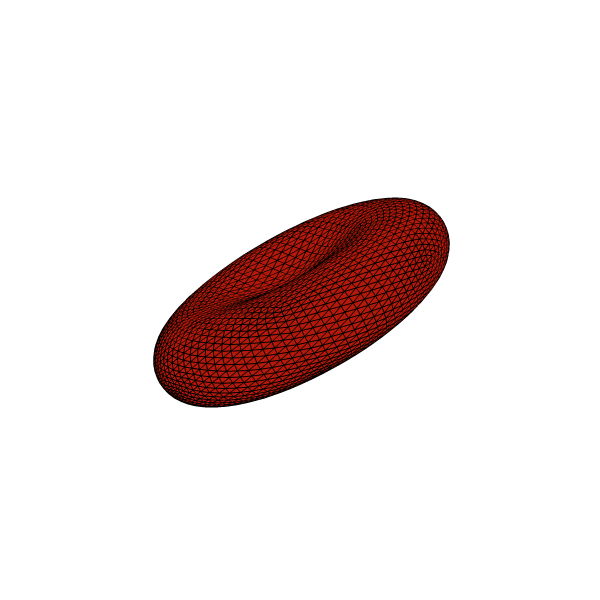}
  \includegraphics[height=.16\columnwidth, trim=130 130 130 130, clip]{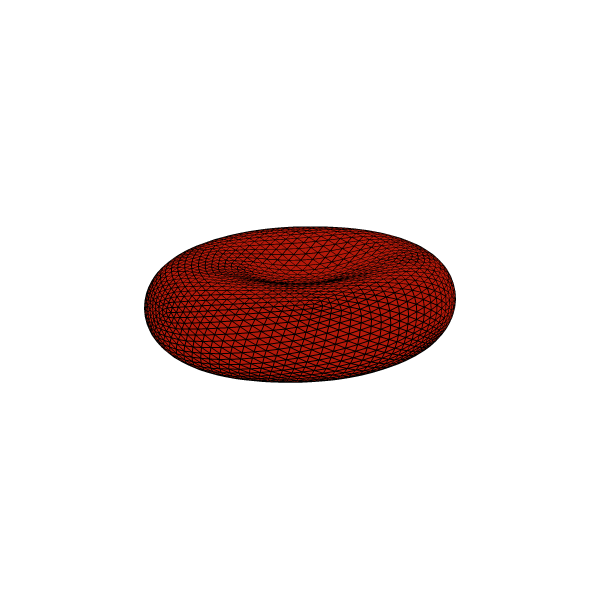}
  \includegraphics[height=.16\columnwidth, trim=150 130 150 130, clip]{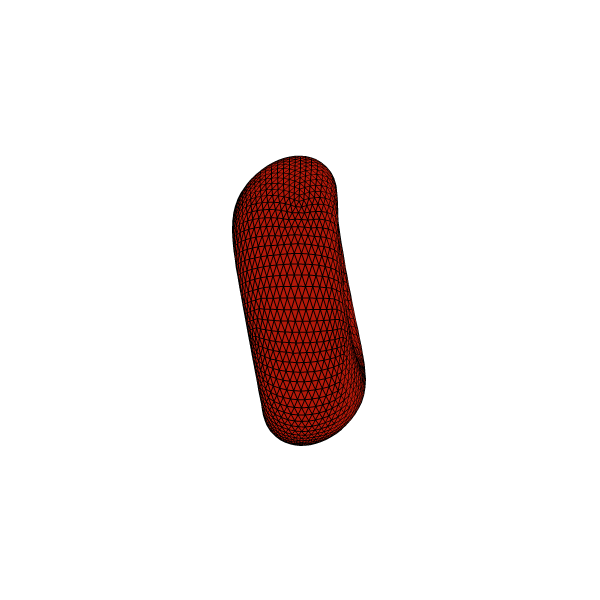}
  \includegraphics[height=.16\columnwidth, trim=130 130 130 130, clip]{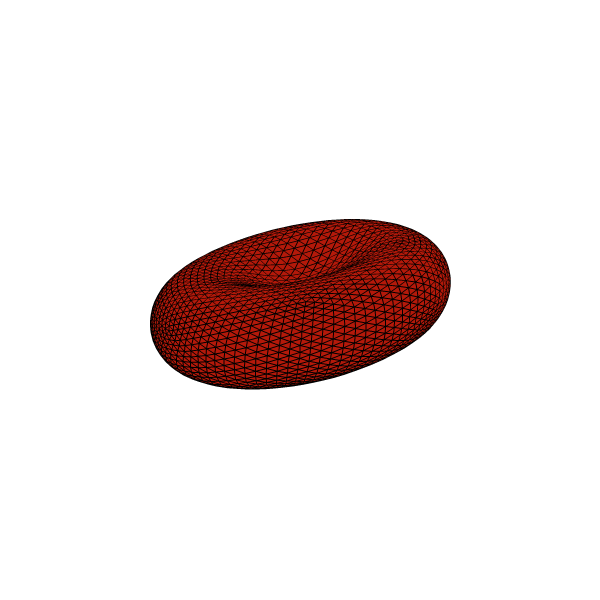}
  \includegraphics[height=.16\columnwidth, trim=150 130 150 130, clip]{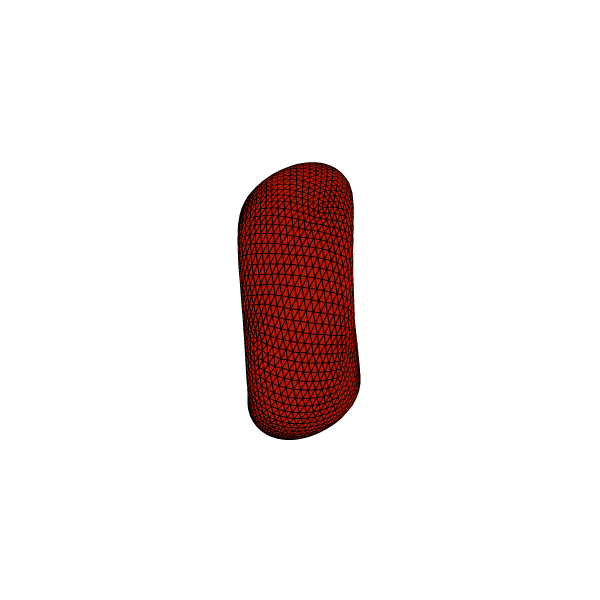}
  \includegraphics[height=.16\columnwidth, trim=130 130 130 130, clip]{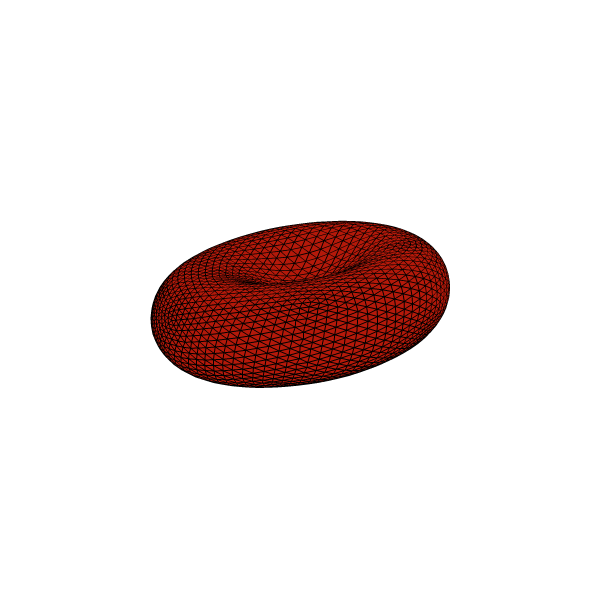}\\
  \includegraphics[width=\columnwidth]{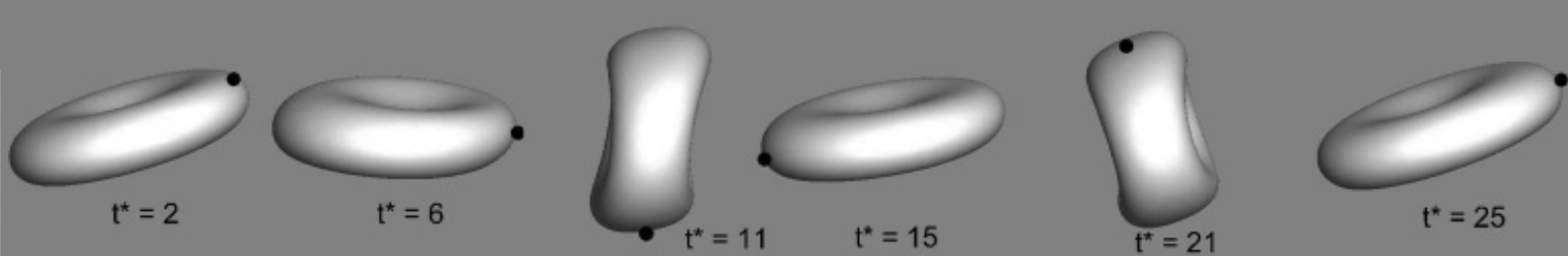}
  \caption{Tumbling motion of an isolated RBC in a shear flow. Top: this study; bottom: Yazdani \& Bagchi \cite{yazdani2011phase}.}
  \label{fig:rbc_shear}
\end{figure}

The qualitative results are presented in \reffig{fig:rbc_shear}, where we include snapshots of the same case from Yazdani \& Bagchi \cite{yazdani2011phase}. We observe that the RBC is undergoing a tumbling motion, a behavior of RBCs that is not seen without viscosity ratio in this range of Capillary numbers \cite{yazdani2011phase, pozrikidis2010computational}. Moreover, the deformation of the RBC matches qualitatively well that of \cite{yazdani2011phase}.
However our tumbling period seems slightly shorter than that of \cite{yazdani2011phase}: we attribute this small discrepancy to the fact that we may have set different values for the area dilatation modulus $C$, as \cite{yazdani2011phase} only provides a range of values: $C \in [50, 400]$.
Nevertheless, the results in \reffig{fig:rbc_shear} show that in our implementation, the combination of elastic forces, bending forces and visocity ratio matches well the qualitative results observed in the literature, and that the overall agreement is satisfactory. The code to reproduce this case is available at \cite{rbc_shear.c}.

\subsection{Capsules interception in a shear flow}
\begin{figure}
  \centering
  \includegraphics[height=.15\textheight]{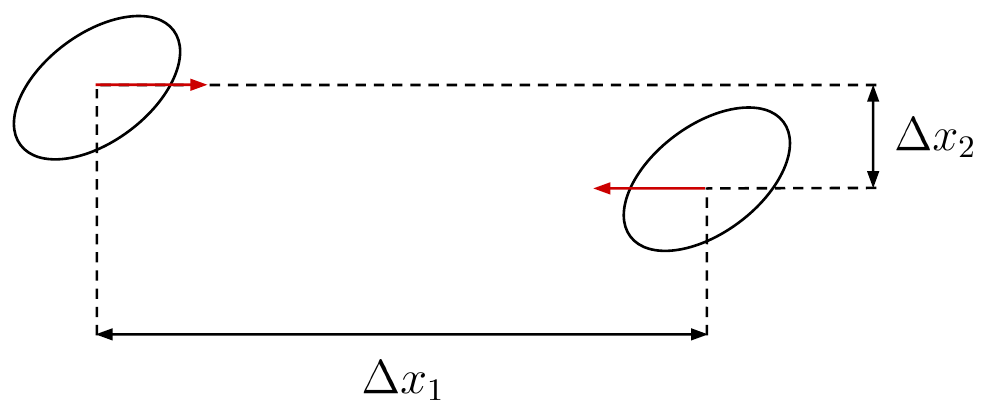}
  \caption{Schematic of the two capsules in in the shear flow, prior to the interception. The horizontal and vertical gaps $\Delta x_1$ and $\Delta x_2$ are defined, and the red arrows represent the velocities of the centers of the capsules.}
  \label{fig:interception_schematic}
\end{figure}

Our last validation case focuses on the interactions of two capsules. Two initially spherical, pre-inflated neo-Hookean capsules are placed in an unbounded shear flow with their initial positions offset in the horizontal and vertical directions as shown in \reffig{fig:interception_schematic}. Since the capsules are offset in the vertical direction, they gain horizontal velocities of opposite signs and their trajectories eventually intercept. This configuration is a good validation candidate since we can compare our results to those obtained by Lac et al. using the boundary integral method \cite{lac2007hydrodynamic}. We consider a computational box of size $16a$ where $a$ is the initial radius of the capsules. The finest Eulerian resolution corresponds to the domain being discretized by $512$ cells in each direction, and the two membranes are discretized with $5120$ triangles. A non-dimensional time step of $\dot{\gamma} \Delta t = 2.5 \cdot 10^{-4}$ is chosen. The Reynolds number is set to $0.01$.

\begin{figure}
  \centering
  \makebox[\textwidth][c]{\includegraphics[height=.7\textheight, trim=135 440 135 120, clip]{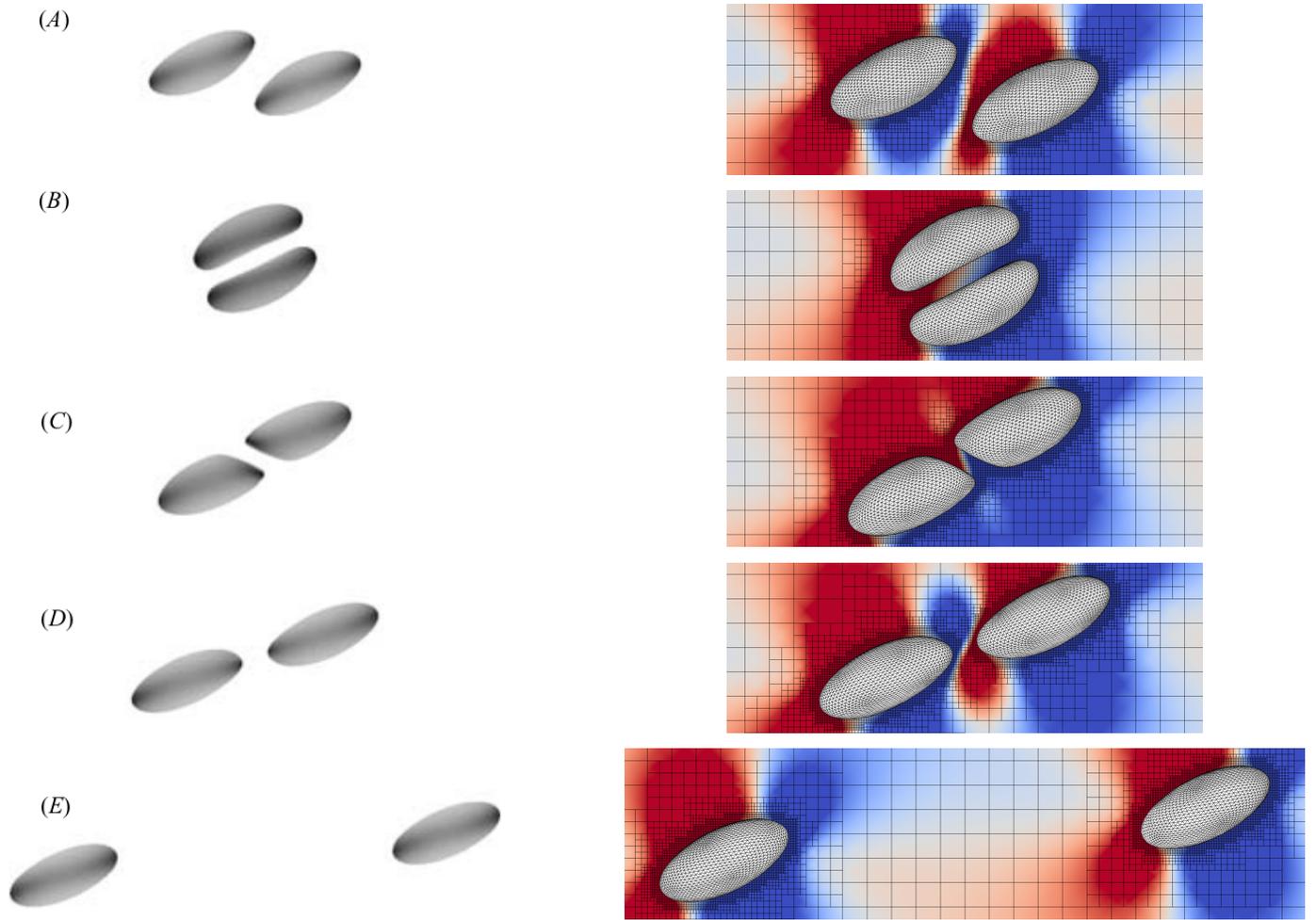}}
  \caption{Snapshots of the interception of two neo-Hookean capsules in a shear flow. Left: Boundary Integral results of Lac et al. \cite{lac2007hydrodynamic}. Right: This study. The color field corresponds to the vertical component of the velocity (rescaled for each snapshot).}
  \label{fig:interception_qualitative}
\end{figure}

\begin{figure}
  \centering
  \makebox[\textwidth][c]{\includegraphics[height=.27\textheight, trim=0 120 0 120, clip]{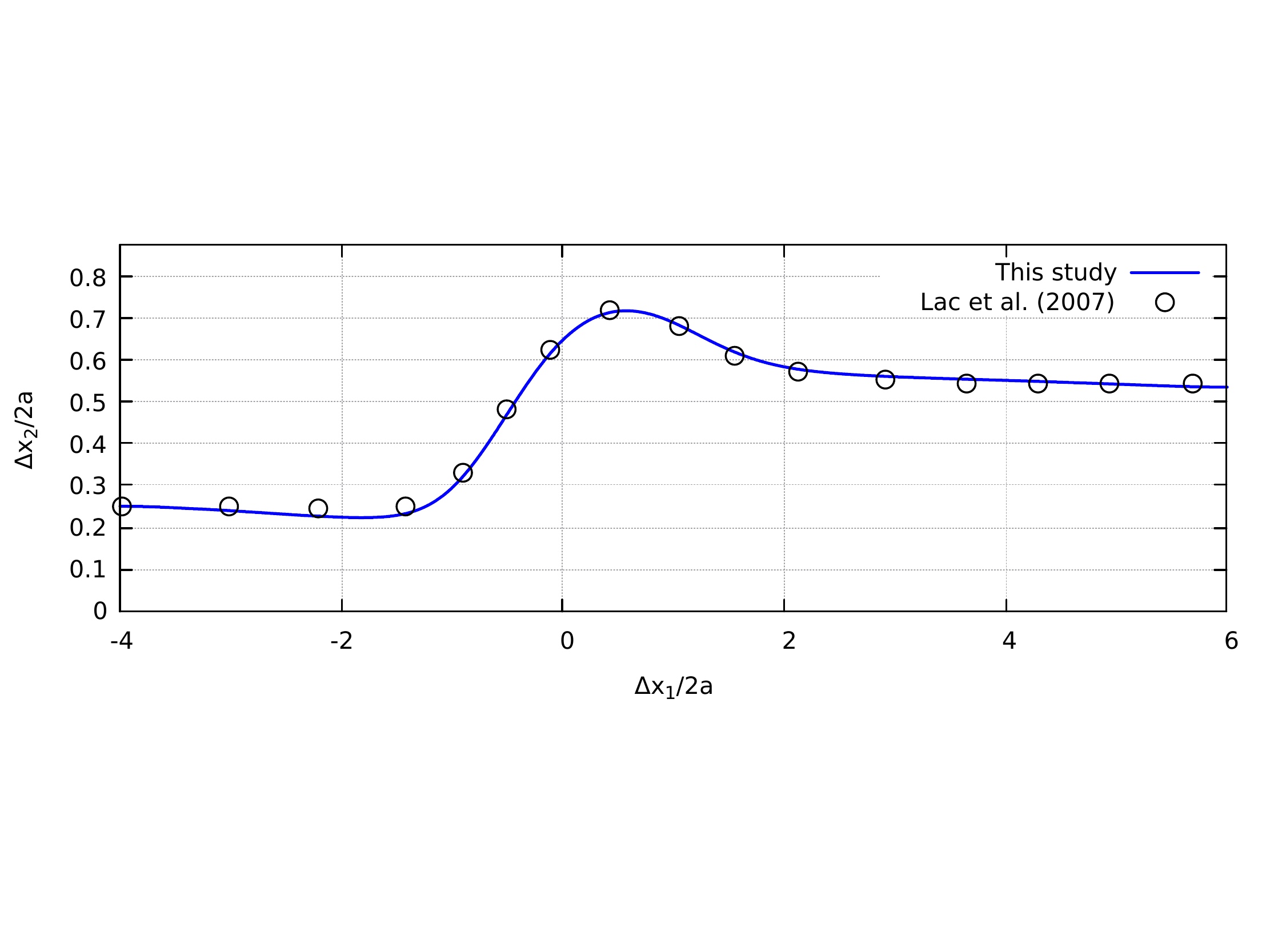}}
  \caption{Non-dimensional vertical gap $\Delta x_2/2a$ against the non-dimension horizontal gap $\Delta x_1/2a$ between the centers of the two capsules. The results from this study are compared to Lac et al. \cite{lac2007hydrodynamic}.}
  \label{fig:interception_quantitative}
\end{figure}

\Reffig{fig:interception_qualitative} shows the qualitative comparison of the shape of the two capsules at several stages of the interception. In our simulations, the color field corresponds to the vertical component of the velocity. At each stage, there is visually no difference in the shape of the capsules. If we track the center of each capsule throughout the simulation, we can compute their difference $\Delta x_2$ in the vertical direction and their difference in the horizontal direction $\Delta x_1$. Normalizing by the initial diameter $2a$ of the capsules, we plot in \reffig{fig:interception_quantitative} the vertical gap between the two capsules as their intercept, and we compare our results to those of Lac et al. \cite{lac2007hydrodynamic}. The agreement is very satisfactory: the transient regime is very well captured, both methods showing a maximum non-dimensional vertical gap of about $0.72$; and the steady-state reached is about $0.54$. Small discrepancies can be observed around $\Delta x_1/2a = -2$ where our vertical gap is slightly lower than that of Lac et al.; and for $\Delta x_1/2a$ between $4$ and $6$ where our slope is still slightly negative while that of Lac et al. is essentially zero.
Those discrepancies are minor and could be explained by our choice of Reynolds number $\text{Re} = 10^{-2}$ while the boundary integral method operates in true Stokes conditions. Regarding the adaptive mesh, as stated above we perform this simulation using an equivalent fluid resolution of $64$ cells per initial diameter, in a cubic box $8$ diameters in length. Our simulation requires about $4.5 \cdot 10^5$ fluid cells, while using a constant mesh size would require about $1.3 \cdot 10^8$ cells. For this specific case, using an adaptive grid therefore reduces the number of fluid cells by a factor of about $300$.

\begin{figure}
  \centering
  \makebox[\textwidth][c]{\includegraphics[height=.4\textheight, trim=0 50 0 50, clip]{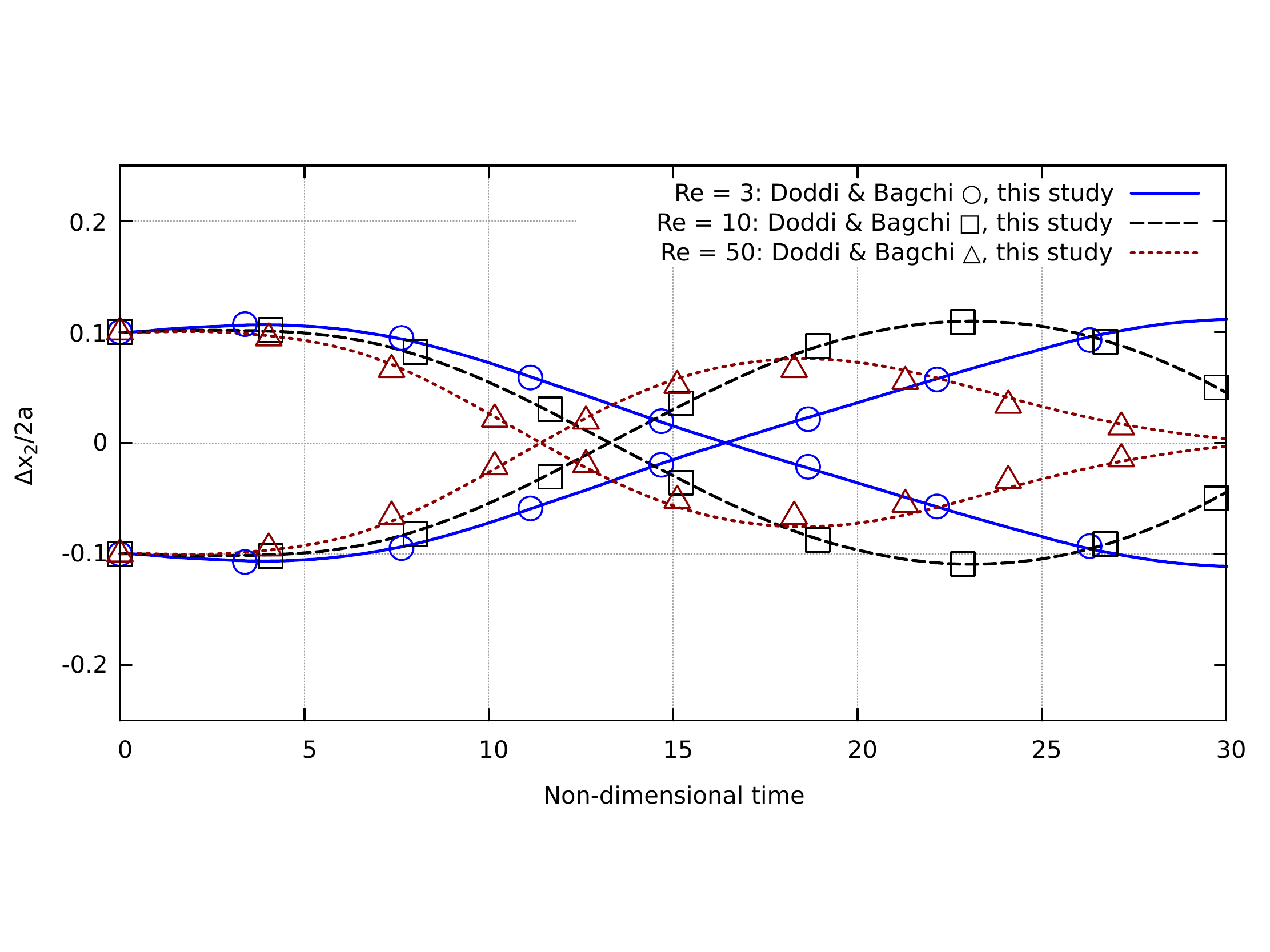}}
  \caption{Non-dimensional vertical gap $\Delta x_2/2a$ against the non-dimensional time $\dot{\gamma} t$ for Reynolds number of 3, 10 and 50. The results of this study are compared to Doddi \& Bagchi \cite{doddi2008effect}.}
  \label{fig:interception_inertia}
\end{figure}

A similar configuration was later examined by Doddi \& Bagchi \cite{doddi2008effect} in the presence of inertia. In a cubic box of size $H = 4 a \pi$, periodic in the $x$ and $z$ directions, for a Capillary number of 0.05 and an initial vertical distance $\Delta x_2/2a = 0.2$, they observed that the capsules don't intercept when the Reynolds number $Re=\rho \dot{\gamma} a^2/\mu$ is greater or equal to 3. Instead, the vertical component of the center of the capsules changes sign and the direction of movement is reversed. We reproduce these results from Doddi \& Bagchi (figure 8c in \cite{doddi2008effect}): each capsule is discretized with $5120$ triangles, the Eulerian equivalent resolution is $40$ points per initial diameter, and the non-dimensional time step is $\Delta t = 10^{-3}$.
\Reffig{fig:interception_inertia} shows the vertical position of the centers of the two capsules with respect to time, for Reynolds numbers of 3, 10 and 50, and the generated data is compared to \cite{doddi2008effect}. Our results superimpose very well with \cite{doddi2008effect}, in particular for $Re = 3$ and $Re = 10$. For $Re = 50$, the agreement is still very satisfactory although we notice that we predict a slightly larger overshoot around $t = 18$ compared to the results of \cite{doddi2008effect}.

Overall, the quantitative agreement in \reffig{fig:interception_quantitative} and \reffig{fig:interception_inertia} is very satisfactory and it validates our adaptive front-tracking solver for several capsules, both in Stokes conditions and in the presence of inertia. The code for these two cases is available at \cite{caps_interception.c, caps_interception_inertia.c}.

\section{Results \label{sec:results}}
\subsection{Capsule flowing through a narrow constriction \label{sec:narrow_constriction}}
In their original study, Park \& Dimitrakopoulos \cite{park2013transient} investigated the case presented in \refsec{sec:constricted_channel} for relatively wide constriction sizes $-$ the half size of the constriction $H_c$ was greater than or equal to the capsule radius $a$. In this subsection we instead decrease the constriction size to $H_c = a/2$ in order to demonstrate the robustness of our solver in cases of extreme deformations, including close to domain boundaries. As in \refsec{sec:constricted_channel}, the capsule is initially circular and pre-inflated such that each distance on the capsule surface is increased by 5\%. It obeys the Skalak elastic law with $C = 1$ and $Ca = 0.1$. Since the capsule undergoes extreme deformations, we also consider a bending force in order to supress unphysically sharp corners: the non-dimensional bending coefficient is set to $\tilde{E_b} = 10^{-3}$. In order to resolve well the capsule deformation, we increase the resolution of the triangulation, which now comprises 20480 triangles. We also perform this simulation for two finest Eulerian grid resolutions of 50 and 100 grid cells per initial diameter respectively.
The flow is driven by an imposed uniform velocity field at the inlet and outlet boundaries, and the Reynolds number is set to 0.01 to model Stokes flow conditions. For this case the non-dimensional time step is set to $a \Delta t /U = 2.5\cdot10^{-5}$, with $a$ the initial radius of the capsule and $U$ the characteristic velocity of the fluid. It appears that this strict restriction on the time step is due to the stiff bending force combined to the very fine discretization we choose.

\begin{figure}
  \centering
  \makebox[\textwidth][c]{\includegraphics[height=.8\textheight, trim=80 150 80 150, clip]{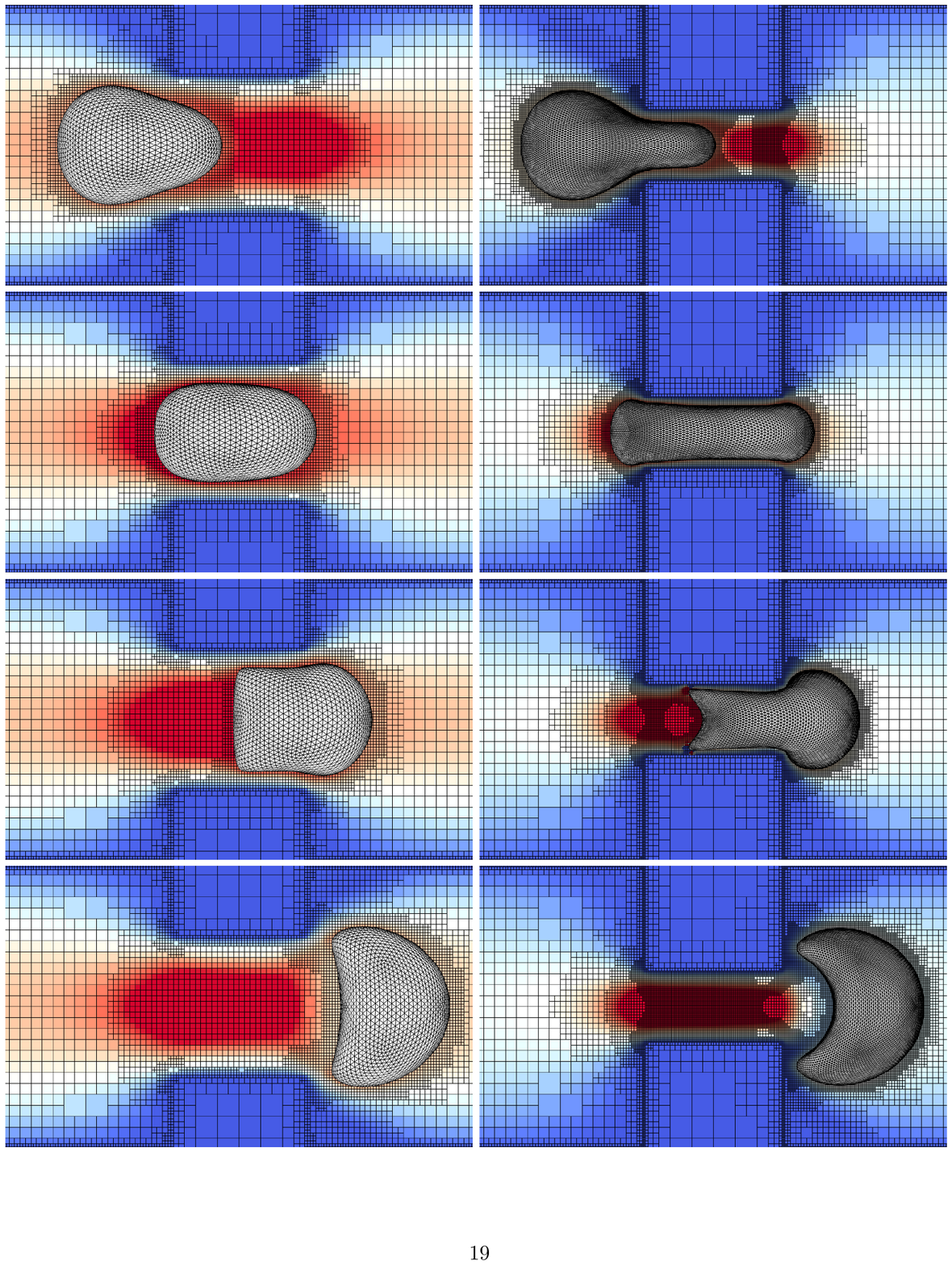}}
  \caption{Qualitative comparison of a capsule flowing through a constricted channel: when the constriction size is equal to the initial diameter of the capsule (left); and when the constriction size is equal to half of the initial diameter of the capsule, with a finest Eulerian grid resolution of 100 cells per initial diameter (right). Color field: $x$-component of the velocity field.}
  \label{fig:narrow_constriction_qualitative}
\end{figure}

\begin{figure}
  \centering
  \includegraphics[height=.43\textheight, trim=0 0 0 35, clip]{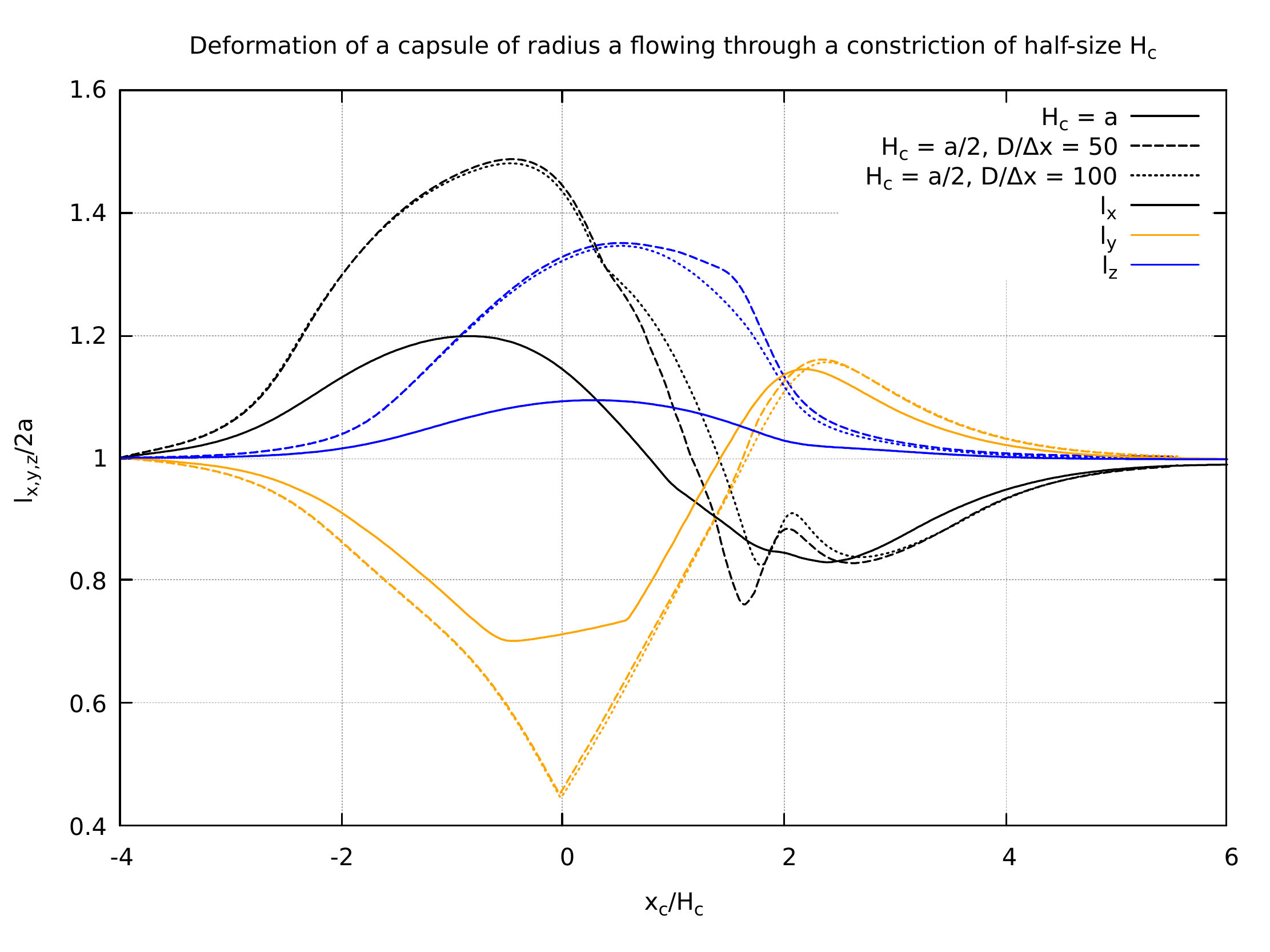}
  \caption{Non-dimensional $x$-, $y$- and $z$-lengths of the capsule $l_x/2a$, $l_y/2a$ and $l_z/2a$ with respect to the non-dimensional $x$-position of the center of the capsule $x_c/H_c$ as it flows though a constriction size equal to: (i) the initial diameter of the capsule (solid line); and (ii) half of the initial diameter of the capsule (dashed line). The dotted line corresponds to a finest Eulerian resolution of 100 Eulerian grid cells per initial diameter (shown for $x_c/H_c = a/2$ only).}
  \label{fig:narrow_constriction_quantitative}
\end{figure}

Qualitative results are shown in \reffig{fig:narrow_constriction_qualitative}. As expected, the deformation of the capsule is considerably greater when the constriction size is halved: the capsule becomes almost flat as it reaches the center of the constriction. \Reffig{fig:narrow_constriction_qualitative} also confirms that the Eulerian mesh is refined only in the region of interest, as the grid size quickly increases with the distance from the constriction and from the capsule. We show quantitative results in \reffig{fig:narrow_constriction_quantitative}, where the non-dimensional $x$-, $y$- and $z$-lengths of the capsule are plotted against the non-dimensional $x$-position of the center of the capsule. Unsurpisingly, the capsule vertical length $l_y$ decreases by over a factor of two when the capsule reaches the center of the constriction, before sharply increasing again as the front of the capsule expands while leaving the constriction. The sharp point observed for $l_y$ at $x_c/H_c = 0$ is simply due to the non-locality of the variable $l_y$: for $x_c/H_c \leq 0$ the maximum height of the capsule is located at its rear, while for $x_c/H_c \geq 0$ it is located at its front. We also note that the maximum decrease of the capsule height $l_{y, \text{min}}^{N}$ in the case of a narrow constriction is much more pronounced than its counterpart $l_{y, \text{min}}^{W}$ in the wider constriction.
However, the minimum capsule height is not halved when the constriction size is halved, i.e. $l_{y, \text{min}}^{N} > l_{y, \text{min}}^{W}/2$ . On the other hand, the maximum $x$-length of the capsule more than doubles in the case of a narrow constriction when compared to the wider constriction size, and the maximum $z$-length more than triples. Therefore, the capsule preference to elongate in the streamwise direction rather than the spanwise direction is reduced when the constriction is narrower. Finally, the capsule reaches a steady shape after the constriction for $x_c > 6 H_c$: the constriction size does not appear to affect this steady shape, which is a slightly deformed sphere compressed in the streamwise direction.

In \reffig{fig:narrow_constriction_quantitative} the $x$-, $y$- and $z$-lengths of the capsule are shown for two maximum Eulerian resolutions: $D/\Delta x = 50$ and $D/\Delta x = 100$, with $D$ the initial diameter of the capsule and $\Delta x$ the smallest Eulerian cell size. Relatively minor differences are observed between the two mesh resolutions, indicating that a maximum Eulerian resolution of $D/\Delta x = 50$ is sufficient to capture the underlying physics of this configuration.
In terms of performance, conducting the previous convergence study up to $D/\Delta x = 100$ with a constant grid size would have required about $4.2\cdot10^7$ fluid cells, while our simulation used less than $4.6 \cdot 10^6$ fluid cells, thus allowing a tenfold reduction in the number of fluid grid cells, and likely reducing the computational resources by a factor of 7 to 10 when accounting for the computational overload due to the complex tree structure of the grid and the adaptivity algorithm.

\subsection{Capsule-laden flows in large complex channel geometries \label{sec:helix}}
\begin{figure}
  \centering
  \begin{minipage}{.53\textwidth}
    \includegraphics[width=\textwidth]{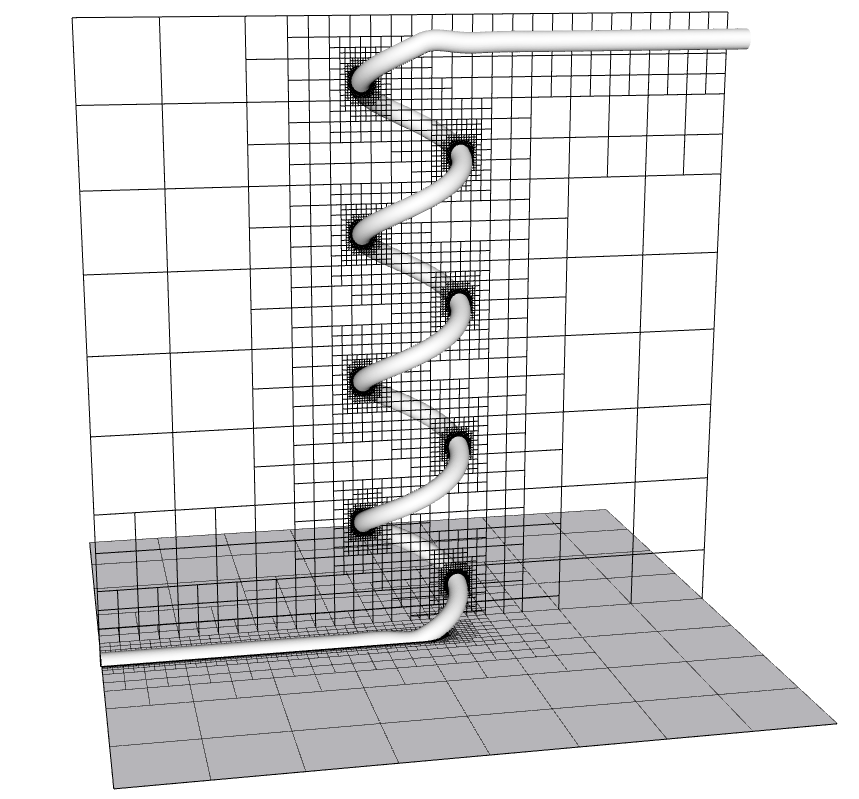}
    \subcaption{}
    \label{fig:helix-geometry}
    \vspace{.5cm}
    \includegraphics[width=\columnwidth]{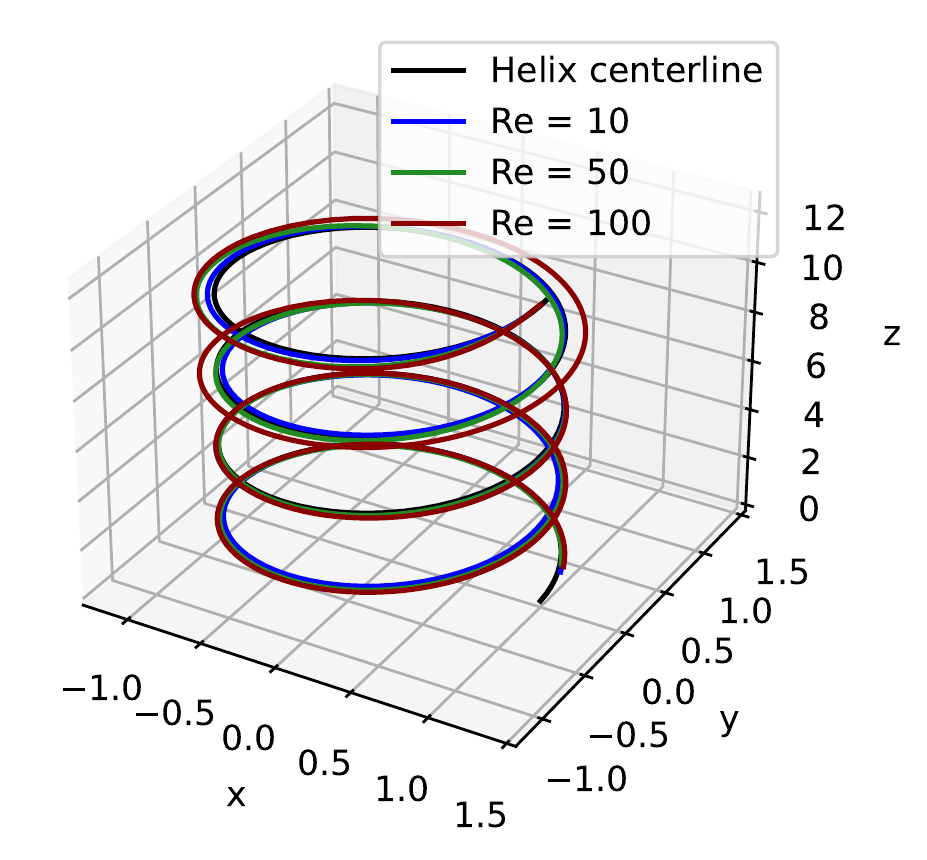}
    \subcaption{}
    \label{fig:helix-3d-path}
  \end{minipage}
  \begin{minipage}{.46\textwidth}
    \includegraphics[width=\textwidth]{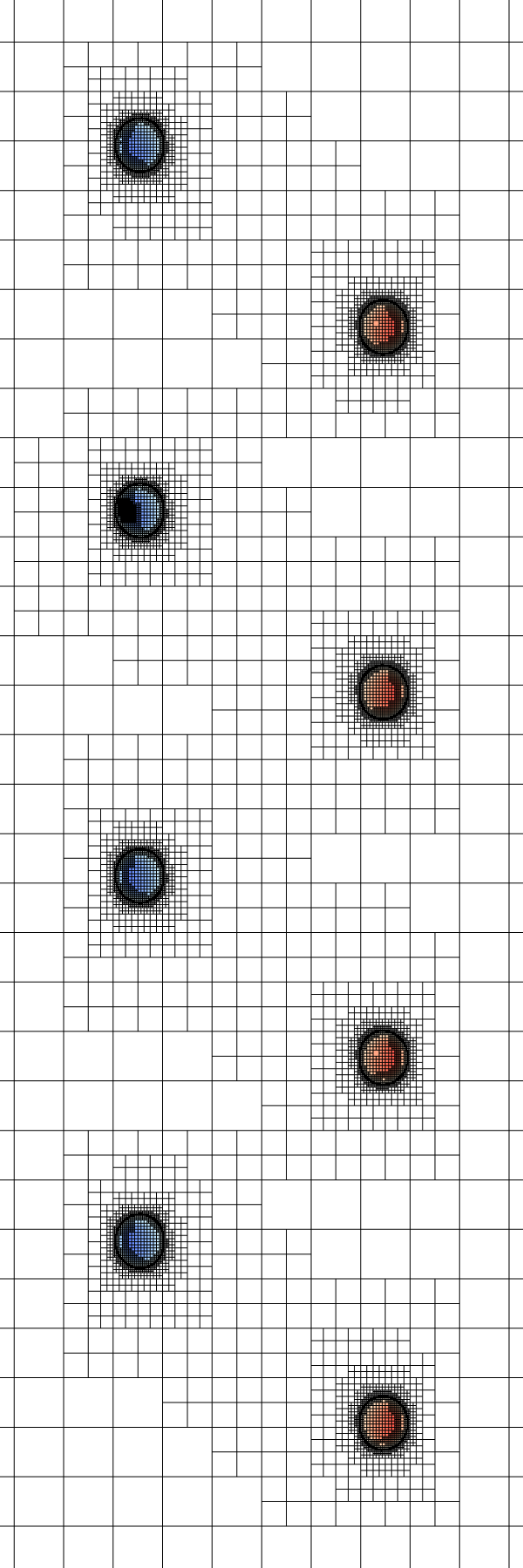}
    \subcaption{}
    \label{fig:helix-ux}
  \end{minipage}
  \caption{Adaptive mesh around a helical geometry and its connecting pipes: (a) full computational box around the whole geometry; (b) three-dimensional trajectory of a capsule at $Re = 10, 50, 100$ (for visual clarity the helix is shrinked in the vertical direction); (c) adaptive mesh and velocity field in the vertical plane. In (c), the color field corresponds to the $x$ component of the velocity, where blue is into the page and orange in out of the page. The capsule is about to cross the vertical plane for the third time, hence the additonal small cells inside the third circular cross-section from the top.}
  \label{fig:helix-mesh}
\end{figure}

It is clear from the previous simulation results presented in this paper that the adaptive mesh refinement is useful to lower the number of cells inside the fluid domain, and thus the amount of computations per time step. However, it can also be desirable to reduce the number of cells \textit{outside} the fluid domain, as they can also be associated with a large computational and memory cost. This is because in cases of complex geometries the computational domain of Cartesian grid methods is by design a bounding box that surrounds the fluid domain. As such, if the volume fraction of the fluid domain in this bounding box is low and if a constant grid size is used, most of the computational cells are located inside the solid walls and a significant amount of memory and computational resources are allocated for these ``solid" cells where no physics happens. This is especially true in cases of large, three-dimensional channel geometries. For instance, let us consider a helical pipe of radius $R_\text{pipe}$ connecting an upper and a lower arm of length $L_\text{arm}$, where a capsule of radius $R_c = R_\text{pipe}/4$ is placed at the top of the geometry. If we assume this geometry is embedded in a bounding box of depth $2R_\text{helix} = 10 R_\text{pipe}$, height $H_\text{helix} = 12 R_\text{helix}$ and length $2 L_\text{arm} = H_\text{helix}$, as shown in \reffig{fig:helix-geometry}, ensuring 16 grid cells per initial capsule diameter using a uniform Eulerian grid would require over 1.2 billion grid cells, rendering the computation extremely expensive. In contrast, using Basilisk's adaptive mesh as shown in \reffig{fig:helix-geometry} allows to reduce the number of computational cells by a factor of about 200, down to less than 6 million grid cells. If only the helix itself is of interest and not the connecting arms, using a uniform Eulerian grid would require about 200 million grid cells, and using our adaptive solver would still reduce the number of grid cells by over a factor of 30.

\begin{figure}
  \centering
  \begin{minipage}{\columnwidth}
    \centering
    \includegraphics[height=.35\textheight]{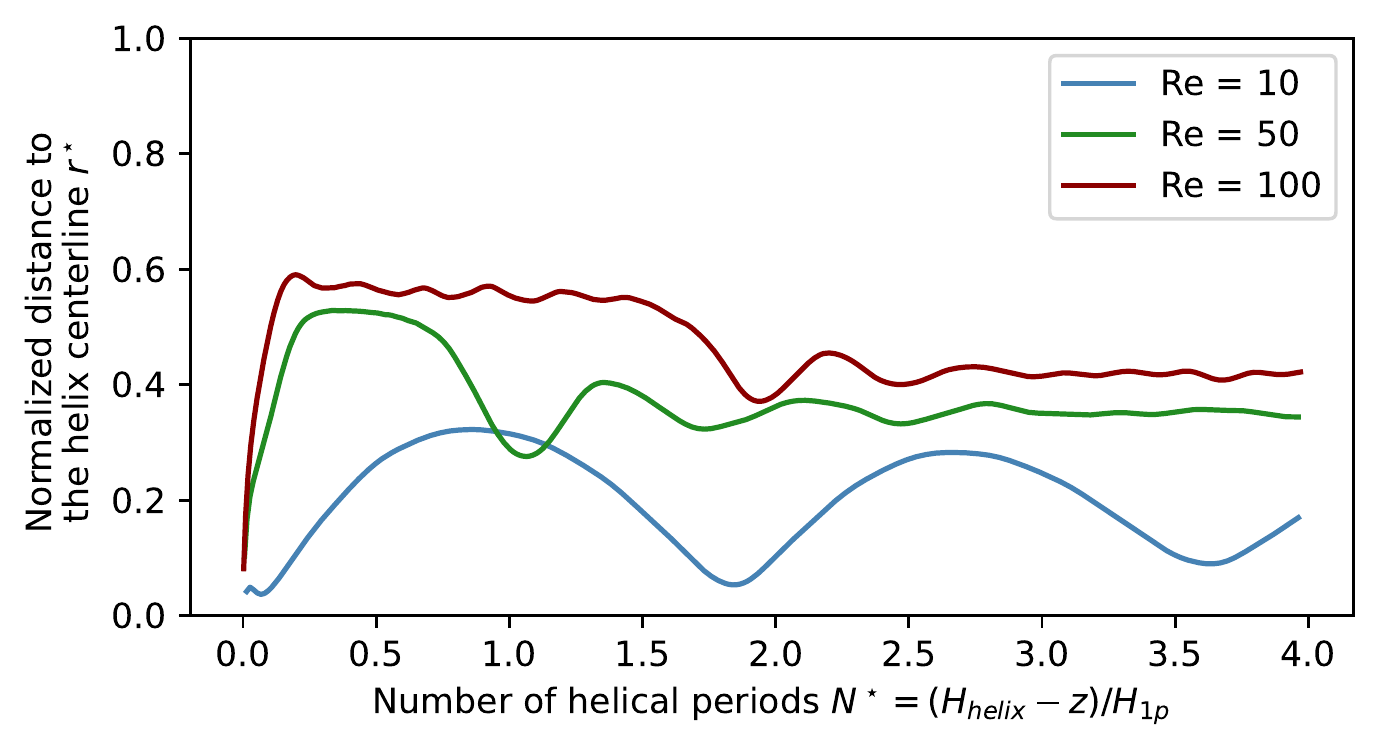}
    \subcaption{}
    \label{fig:helix-radial-migration}
  \end{minipage}\\
  \begin{minipage}{\columnwidth}
    \centering
    \includegraphics[height=.35\textheight, trim=0 30 0 0, clip]{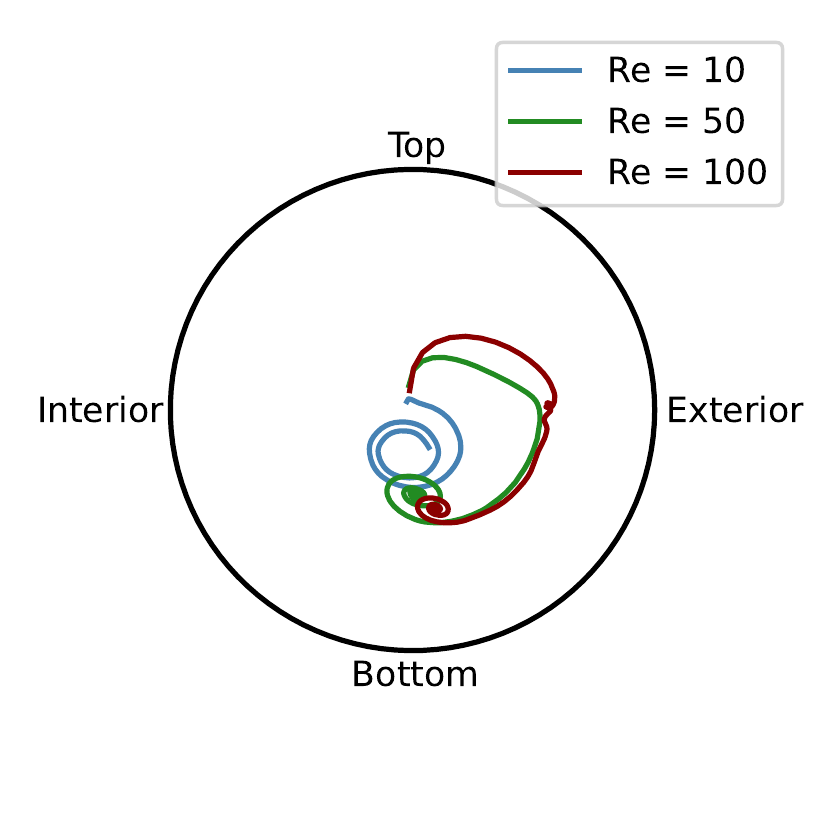}
    \subcaption{}
    \label{fig:helix-plane-migration}
  \end{minipage}
  \caption{Trajectory of a solitary capsule in the helix at $Re=10, 50, 100$: (a) normalized distance from the helix centerline; (b) radial migration in a plane orthogonal to the helix centerline. $N^\star$ corresponds to the number of revolutions around the vertical axis of the helix.}
  \label{fig:helix-quantitative}
\end{figure}

As a demonstration of the capability of the present solver to handle such large and complex geometries, simulations are carried out in the helical geometry described above and shown in \reffig{fig:helix-mesh}. A neo-Hookean capsule of radius $R_c$ with $Ca = 0.1$ is placed on the upper straight pipe centerline at a distance of $1.5R_\text{helix}$ prior to entering the helix. We propose to study the inertial migration of this capsule for three Reynolds numbers: $Re = 10, 50$ and $100$. We set the non-dimensional time step $R_\text{helix}\Delta t / U$ to $10^{-3}$, with $U$ the characteristic velocity of the fluid; and we choose a finest Eulerian grid resolution corresponding to 16 grid cells per initial diameter. Each case ran for about three days on 96 processors, with around $6 \cdot 10^4$ cells per processor. To analyze the trajectories, we define the distance $r^\star$ of the capsule centroid to the helix centerline normalized by the pipe radius $R_\text{pipe}$, as well as the number of helical periods $N^\star = (H_\text{helix} - z)/H_{1p}$, with $H_{1p}$ the height of one vertical period of the helix, i.e. its pitch distance.
A slice of the flow field in the helix for $Re = 100$ is shown in \reffig{fig:helix-ux}, and a movie of this simulation as well as the code to reproduce it are available at \cite{helix.c}. The trajectories of the capsule are shown in the three-dimensional space in \reffig{fig:helix-3d-path}, in one dimension by showing $r^\star$ as a function of $N^\star$ in \reffig{fig:helix-radial-migration}, and in a cross-section of the pipe orthogonal to the helix centerline in \reffig{fig:helix-plane-migration}. Only the path corresponding to the capsule located inside the helix is shown in these figures. Immediately after release, the capsule moves away from the centerline for all Reynolds numbers. The initial overshoot of $r^\star$ increases with the Reynolds number, from $r^\star_\text{max} \approx 0.3$ at $Re = 10$ to $r^\star_\text{max} \approx 0.6$ at $Re = 100$.
After four helical revolutions, the capsule exits the helix with a steady position of $r^\star_\infty \approx 0.38$ and $r^\star_\infty \approx 0.45$ for $Re = 50$ and $Re = 100$ respectively. In the case of $Re = 10$, however, a steady state is not yet reached when the capsule exits the helix, but we can extrapolate the capsule trajectory to find that $r^\star_\infty \approx 0.18$ for this Reynolds number. For all Reynolds numbers, the steady position is located in the lower half of the cross-section, at an angle $\theta$ from the horizontal line of about $-\pi/2$, although $\theta$ increases slightly with the Reynolds number.
Interestingly, we note that the capsule transient path is longer for $Re = 100$ than that for $Re = 50$: as can be seen in \reffig{fig:helix-quantitative}, for $Re = 100$ the capsule seems to reach an unstable equilibrium at $r^\star \approx 0.58$ and $\theta \approx 0$ for as long as 1.5 helical periods, before continuing its spiralling motion towards a stable steady-state location. Further investigation would be necessary to characterize this behavior and determine if, for instance, this unstable equilibrium corresponds to the center of a vortex, but this is not the focus of the present paper. The purpose of this simulation is to show that the present solver is able to simulate large three-dimensional channel geometries, and has the potential to simulate full microfluidic devices.

\section{Conclusion and perspectives\label{sec:conclusion}}
We have presented an adaptive front-tracking solver to simulate deformable capsules in viscous flows at zero and finite inertia. The membrane mechanics is governed by an elastic and a bending resistance, and a non-unity viscosity ratio is allowed. Moreover, the present solver is compatible with complex STL-defined geometries, thus providing all the ingredients needed to simulate realistic flow configurations of biological cells, including red blood cells, both in-vivo and in-vitro. Numerous validation cases are presented against data available in the literature: we compare our results mainly to the highly accurate boundary integral method in Stokes conditions, and to other front-tracking methods at non zero Reynolds numbers. Very good qualitative and quantitative agreement is shown in all cases. We then demonstrate the robustness of the present solver in more challenging configurations, as well as its potential to tackle very large, three-dimensional channel geometries relevant to inertial microfluidic applications.
Moreover, the present implementation is open-source as part of the Basilisk platfrom: the documented source code is freely available, as well as the source files to reproduce all the simulations presented in this paper \cite{huet_sandbox}.

Although the present adaptive front-tracking solver can simulate all the range of Reynolds numbers, the non-inertial limit is challenging because of the computation of the viscous term. The simulations we show in this paper at $Re = 0.01$ are several times slower to complete than their counterpart at, e.g., $Re = O(1)$ or $Re = O(10)$. As a result, if only the non-inertial regime is sought, boundary integral solvers likely remain the most efficient method by far. Another challenge is the stiffness of the bending stresses: since the Helfrich's bending formulation involves such high-order derivatives of the membrane geometry, and since the time integration of the membrane problem is explicit, the time step is controlled by the time scale associated with the bending force whenever bending effects are included. This is a known challenge of computing bending stresses on elastic membranes (see, e.g., p. 40 of \cite{barthes2016motion}). As a result, we have to decrease our time step by one order of magnitude (sometimes even more) whenever the bending stresses are included. Unfortunately, to our knowledge the stability condition associated with the bending force is unknown and it is therefore not possible to stabilize simulations by employing an adaptive time-step strategy, as is already the case in Basilisk with the CFL condition and with the celerity of capillary waves for surface tension stresses. One could investigate the implicit or ``approximately implicit" treatment of the immersed boundary method as done by Roma, Peskin \& Berger \cite{Roma1999}.

On the implementation side, the fluid solver from Basilisk is compatible with shared and distributed memory systems $-$ i.e. using OpenMP and MPI libraries $-$, allowing to run simulations on large supercomputers. Naturally, we have enabled our front-tracking solver to be compatible with MPI as well. However, ensuring a good scaling with the number of capsules is not trivial as the domain decomposition of the Eulerian adaptive mesh is governed by a Z-ordering algorithm. As a result, when the Eulerian mesh is adaptive, the stencils attached to the Lagrangian nodes of a given capsule are most likely containing Eulerian cells handled by several distinct processors. In other words, a single capsule has to exist on many different processors in order to communicate with the background Eulerian mesh. Interpolating velocities from the fluid cells to the Lagrangian nodes, and spreading the Lagrangian forces to the fluid cells thus requires expensive inter-processor communications. Investigating efficient strategies to simulate a large number of capsules with an adaptive mesh is left for future works. That being said, in all the simulations shown in this study with one or two capsules considered, at most $5\%$ of the total computation time is spent in the front-tracking solver. Consequently, unless simulating a large number of capsules $-$ e.g., $O(100)$ $-$, the bottleneck is still the Navier-Stokes solver. Moreover when a large number of capsules is considered, one could argue that a uniform Eulerian mesh can be more efficient than its adaptive counterpart because a large number of capsules would likely result in a high volume fraction of capsules. An efficient parallel implementation of the present front-tracking solver restricted to uniform Eulerian grids is straightforward and could be implemented in future studies if dense volume fractions are considered. Another extension to the present solver could be to allow the triangulation of the membrane to be adaptive as well. Such adaptive triangulations have been considered to simulate fluid-fluid interfaces \cite{tryggvason2001front}, but in the case of elastic membranes special care needs to be given to coarsen or refine the shape functions of a triangle. However only simulations featuring extreme membrane deformation would benefit from an adaptive membrane triangulation, as the front-tracking solver is only taking a few percents of the total computing time. For the applications we seek where reasonable membrane deformations are expected, the gain of an adaptive membrane triangulation is likely close to zero.

Another possible improvement to the present solver would be to allow the support of the regularized Dirac-delta functions to include grid cells of different sizes. Indeed, the current method imposes a constant grid size in the vicinity of the membrane in order to apply the IBM in a straightforward manner, in a similar fashion to our Distributed Lagrange Multriplier/Fictitious Domain method implemented in Basilisk to simulate flows laden with rigid particles \cite{selccuk2021fictitious}, but this can result in imposing a finer Eulerian grid resolution around the membrane than what is necessary to properly resolve some parts of the membrane. This scenario typically happens when the capsule is close or will come close to a sharp boundary, such as in the case of the narrow constriction in \refsec{sec:narrow_constriction}. In \reffig{fig:narrow_constriction_qualitative}, for instance, as the capsule enters the narrow constriction, the Eulerian grid resolution around its tail is much finer than necessary due to the very fine grid resolution needed at the front of the capsule, which is located in a flow with strong gradients and close to sharp boundaries. A conceptually simple way to allow the size of the Eulerian grid cells to change along the membrane would be: (i) to propagate the forcing term or the averaging operator to smaller grid cells located inside the stencil of interest; and (ii) to increase the stencil size if a large grid cell is encountered, such that any stencil size is always four times larger in each direction than its largest grid cell. This adaptive extension of the immersed boundary method is under current investigation and shows promising results.

As demonstrated in \refsec{sec:helix}, the current state of our adaptive solver allows resolved inertial simulations of capsule-laden flows in large three-dimensional geometries for a fraction of the computational cost of that of uniform front-tracking solvers implemented on uniform Cartesian grids. As such, our solver has the potential to provide valuable insight to help develop inertial migration microfluidic devices. The present solver may even allow the simulation of full microfluidic geometries consisting in several stacked layers of microfluidic channels, such as the spiralling geometries in the experimental work of Fang et al. \cite{fang2022efficient}. This has the potential to provide valuable qualitative and quantitative information about the flow field, capsule dynamics and sorting efficiency of a given realistic microfluidic geometry, thus reducing the number of manufacturing iterations during the design process of such devices. Our medium-term objective is to tackle this type of virtual design problem, considering sub-domains of the full geometry as a first step, while concomitently investigating the possible improvements stated above.

\section*{Acknowledgements}
Damien P. Huet thanks Antoine Morente for designing the helix STL geometry and for helpful discussions. The authors greatly appreciate the ﬁnancial support of the Natural Sciences and Engineering Research Council of Canada (NSERC)
via Anthony Wachs’ Discovery Grant RGPIN-2016-06572. This research was enabled by support provided by Compute Canada (\url{http://www.computecanada.ca})
through Anthony Wachs’s 2020 and 2021 Computing Resources for Research Groups allocation qpf-764-ac.

\bibliographystyle{elsarticle/elsarticle-num}
\bibliography{references}

\begin{thebibliography}{10}
\expandafter\ifx\csname url\endcsname\relax
  \def\url#1{\texttt{#1}}\fi
\expandafter\ifx\csname urlprefix\endcsname\relax\def\urlprefix{URL }\fi
\expandafter\ifx\csname href\endcsname\relax
  \def\href#1#2{#2} \def\path#1{#1}\fi

\bibitem{dewhirst2017transport}
M.~W. Dewhirst, T.~W. Secomb, Transport of drugs from blood vessels to tumour
  tissue, Nature Reviews Cancer 17~(12) (2017) 738--750.

\bibitem{puleri2021computational}
D.~F. Puleri, P.~Balogh, A.~Randles, Computational models of cancer cell
  transport through the microcirculation, Biomechanics and Modeling in
  Mechanobiology 20~(4) (2021) 1209--1230.

\bibitem{balogh2021data}
P.~Balogh, J.~Gounley, S.~Roychowdhury, A.~Randles, A data-driven approach to
  modeling cancer cell mechanics during microcirculatory transport, Scientific
  Reports 11~(1) (2021) 1--18.

\bibitem{islamzada2020deformability}
E.~Islamzada, K.~Matthews, Q.~Guo, A.~T. Santoso, S.~P. Duffy, M.~D. Scott,
  H.~Ma, Deformability based sorting of stored red blood cells reveals
  donor-dependent aging curves, Lab on a Chip 20~(2) (2020) 226--235.

\bibitem{bazaz2022zigzag}
S.~R. Bazaz, A.~Mihandust, R.~Salomon, H.~A.~N. Joushani, W.~Li, H.~A. Amiri,
  F.~Mirakhorli, S.~Zhand, J.~Shrestha, M.~Miansari, et~al., Zigzag
  microchannel for rigid inertial separation and enrichment (z-rise) of cells
  and particles, Lab on a Chip.

\bibitem{takeishi2022inertial}
N.~Takeishi, H.~Yamashita, T.~Omori, N.~Yokoyama, S.~Wada, M.~Sugihara-Seki,
  Inertial migration of red blood cells under a newtonian fluid in a circular
  channel, arXiv preprint arXiv:2209.11933.

\bibitem{gangadhar2022inertial}
A.~Gangadhar, S.~A. Vanapalli, Inertial focusing of particles and cells in the
  microfluidic labyrinth device: Role of sharp turns, Biomicrofluidics 16~(4)
  (2022) 044114.

\bibitem{fang2022efficient}
Y.~Fang, S.~Zhu, W.~Cheng, Z.~Ni, N.~Xiang, Efficient bioparticle extraction
  using a miniaturized inertial microfluidic centrifuge, Lab on a Chip 22~(18)
  (2022) 3545--3554.

\bibitem{barthes1981time}
D.~Barthes-Biesel, J.~Rallison, The time-dependent deformation of a capsule
  freely suspended in a linear shear flow, Journal of Fluid Mechanics 113
  (1981) 251--267.

\bibitem{pozrikidis1995finite}
C.~Pozrikidis, Finite deformation of liquid capsules enclosed by elastic
  membranes in simple shear flow, Journal of Fluid Mechanics 297 (1995)
  123--152.

\bibitem{ramanujan1998deformation}
S.~Ramanujan, C.~Pozrikidis, Deformation of liquid capsules enclosed by elastic
  membranes in simple shear flow: large deformations and the effect of fluid
  viscosities, Journal of fluid mechanics 361 (1998) 117--143.

\bibitem{eggleton1998large}
C.~D. Eggleton, A.~S. Popel, Large deformation of red blood cell ghosts in a
  simple shear flow, Physics of fluids 10~(8) (1998) 1834--1845.

\bibitem{pozrikidis2001effect}
C.~Pozrikidis, Effect of membrane bending stiffness on the deformation of
  capsules in simple shear flow, Journal of Fluid Mechanics 440 (2001) 269.

\bibitem{pozrikidis2003numerical}
C.~Pozrikidis, Numerical simulation of the flow-induced deformation of red
  blood cells, Annals of biomedical engineering 31~(10) (2003) 1194--1205.

\bibitem{zhao2010spectral}
H.~Zhao, A.~H. Isfahani, L.~N. Olson, J.~B. Freund, A spectral boundary
  integral method for flowing blood cells, Journal of Computational Physics
  229~(10) (2010) 3726--3744.

\bibitem{lac2004spherical}
E.~Lac, D.~Barthes-Biesel, N.~Pelekasis, J.~Tsamopoulos, Spherical capsules in
  three-dimensional unbounded stokes flows: effect of the membrane constitutive
  law and onset of buckling, Journal of Fluid Mechanics 516 (2004) 303--334.

\bibitem{lac2007hydrodynamic}
E.~Lac, A.~Morel, D.~Barth{\`e}s-Biesel, Hydrodynamic interaction between two
  identical capsules in simple shear flow, Journal of Fluid Mechanics 573
  (2007) 149--169.

\bibitem{bagchi2007mesoscale}
P.~Bagchi, Mesoscale simulation of blood flow in small vessels, Biophysical
  journal 92~(6) (2007) 1858--1877.

\bibitem{doddi2008lateral}
S.~K. Doddi, P.~Bagchi, Lateral migration of a capsule in a plane poiseuille
  flow in a channel, International Journal of Multiphase Flow 34~(10) (2008)
  966--986.

\bibitem{yazdani2011phase}
A.~Z. Yazdani, P.~Bagchi, Phase diagram and breathing dynamics of a single red
  blood cell and a biconcave capsule in dilute shear flow, Physical Review E
  84~(2) (2011) 026314.

\bibitem{yazdani2012three}
A.~Yazdani, P.~Bagchi, Three-dimensional numerical simulation of vesicle
  dynamics using a front-tracking method, Physical Review E 85~(5) (2012)
  056308.

\bibitem{yazdani2013influence}
A.~Yazdani, P.~Bagchi, Influence of membrane viscosity on capsule dynamics in
  shear flow, Journal of fluid mechanics 718 (2013) 569--595.

\bibitem{doddi2008effect}
S.~K. Doddi, P.~Bagchi, Effect of inertia on the hydrodynamic interaction
  between two liquid capsules in simple shear flow, International journal of
  multiphase flow 34~(4) (2008) 375--392.

\bibitem{balogh2017computational}
P.~Balogh, P.~Bagchi, A computational approach to modeling cellular-scale blood
  flow in complex geometry, Journal of Computational Physics 334 (2017)
  280--307.

\bibitem{balogh2018analysis}
P.~Balogh, P.~Bagchi, Analysis of red blood cell partitioning at bifurcations
  in simulated microvascular networks, Physics of Fluids 30~(5) (2018) 051902.

\bibitem{li2008front}
X.~Li, K.~Sarkar, Front tracking simulation of deformation and buckling
  instability of a liquid capsule enclosed by an elastic membrane, Journal of
  Computational Physics 227~(10) (2008) 4998--5018.

\bibitem{zhang2007immersed}
J.~Zhang, P.~C. Johnson, A.~S. Popel, An immersed boundary lattice boltzmann
  approach to simulate deformable liquid capsules and its application to
  microscopic blood flows, Physical biology 4~(4) (2007) 285.

\bibitem{zhang2008red}
J.~Zhang, P.~C. Johnson, A.~S. Popel, Red blood cell aggregation and
  dissociation in shear flows simulated by lattice boltzmann method, Journal of
  biomechanics 41~(1) (2008) 47--55.

\bibitem{ames2020multi}
J.~Ames, D.~F. Puleri, P.~Balogh, J.~Gounley, E.~W. Draeger, A.~Randles,
  Multi-gpu immersed boundary method hemodynamics simulations, Journal of
  computational science 44 (2020) 101153.

\bibitem{fedosov2010multiscale}
D.~A. Fedosov, B.~Caswell, G.~E. Karniadakis, A multiscale red blood cell model
  with accurate mechanics, rheology, and dynamics, Biophysical journal 98~(10)
  (2010) 2215--2225.

\bibitem{hirt1981volume}
C.~W. Hirt, B.~D. Nichols, Volume of fluid (vof) method for the dynamics of
  free boundaries, Journal of computational physics 39~(1) (1981) 201--225.

\bibitem{brackbill1992continuum}
J.~U. Brackbill, D.~B. Kothe, C.~Zemach, A continuum method for modeling
  surface tension, Journal of computational physics 100~(2) (1992) 335--354.

\bibitem{tryggvason2001front}
G.~Tryggvason, B.~Bunner, A.~Esmaeeli, D.~Juric, N.~Al-Rawahi, W.~Tauber,
  J.~Han, S.~Nas, Y.-J. Jan, A front-tracking method for the computations of
  multiphase flow, Journal of computational physics 169~(2) (2001) 708--759.

\bibitem{popinet2009accurate}
S.~Popinet, An accurate adaptive solver for surface-tension-driven interfacial
  flows, Journal of Computational Physics 228~(16) (2009) 5838--5866.

\bibitem{cottet2006level}
G.-H. Cottet, E.~Maitre, A level set method for fluid-structure interactions
  with immersed surfaces, Mathematical models and methods in applied sciences
  16~(03) (2006) 415--438.

\bibitem{ii2012full}
S.~Ii, X.~Gong, K.~Sugiyama, J.~Wu, H.~Huang, S.~Takagi, A full eulerian
  fluid-membrane coupling method with a smoothed volume-of-fluid approach,
  Communications in Computational Physics 12~(2) (2012) 544.

\bibitem{ii2012computational}
S.~Ii, K.~Sugiyama, S.~Takagi, Y.~Matsumoto, A computational blood flow
  analysis in a capillary vessel including multiple red blood cells and
  platelets, Journal of Biomechanical Science and Engineering 7~(1) (2012)
  72--83.

\bibitem{ii2018continuum}
S.~Ii, K.~Shimizu, K.~Sugiyama, S.~Takagi, Continuum and stochastic approach
  for cell adhesion process based on eulerian fluid-capsule coupling with
  lagrangian markers, Journal of Computational Physics 374 (2018) 769--786.

\bibitem{Uhlmann2005}
M.~Uhlmann, {An immersed boundary method with direct forcing for the simulation
  of particulate flows}, Journal of Computational Physics 209~(2) (2005)
  448--476.

\bibitem{Kempe2012a}
T.~Kempe, J.~Fr{\"o}hlich, {An improved immersed boundary method with direct
  forcing for the simulation of particle laden flows}, Journal of Computational
  Physics 231~(9) (2012) 3663--3684.

\bibitem{breugem2012second}
W.-P. Breugem, A second-order accurate immersed boundary method for fully
  resolved simulations of particle-laden flows, Journal of Computational
  Physics 231~(13) (2012) 4469--4498.

\bibitem{Roma1999}
A.~Roma, C.~Peskin, M.~Berger, {An adaptive version of the immersed boundary
  method}, Journal of Computational Physics 153~(2) (1999) 509--534.

\bibitem{griffith2007adaptive}
B.~E. Griffith, R.~D. Hornung, D.~M. McQueen, C.~S. Peskin, An adaptive,
  formally second order accurate version of the immersed boundary method,
  Journal of computational physics 223~(1) (2007) 10--49.

\bibitem{vanella2014adaptive}
M.~Vanella, A.~Posa, E.~Balaras, Adaptive mesh refinement for immersed boundary
  methods, Journal of Fluids Engineering 136~(4).

\bibitem{agresar1998adaptive}
G.~Agresar, J.~Linderman, G.~Tryggvason, K.~Powell, An adaptive, cartesian,
  front-tracking method for the motion, deformation and adhesion of circulating
  cells, Journal of Computational Physics 143~(2) (1998) 346--380.

\bibitem{cheng2022immersed}
Z.~Cheng, A.~Wachs, An immersed boundary/multi-relaxation time lattice
  boltzmann method on adaptive octree grids for the particle-resolved
  simulation of particle-laden flows, Journal of Computational Physics (2022)
  111669.

\bibitem{popinet2003gerris}
S.~Popinet, Gerris: a tree-based adaptive solver for the incompressible euler
  equations in complex geometries, Journal of Computational Physics 190~(2)
  (2003) 572--600.

\bibitem{Popinet2015}
S.~Popinet, {A quadtree-adaptive multigrid solver for the Serre--Green--Naghdi
  equations}, Journal of Computational Physics 302 (2015) 336--358.

\bibitem{huet_sandbox}
D.~P. Huet, \url{http://basilisk.fr/sandbox/huet}, accessed: 2022-10-05 (2022).

\bibitem{green1960large}
A.~E. Green, J.~E. Adkins, Large elastic deformations and non-linear continuum
  mechanics.

\bibitem{barthes2002effect}
D.~Barthes-Biesel, A.~Diaz, E.~Dhenin, Effect of constitutive laws for
  two-dimensional membranes on flow-induced capsule deformation, Journal of
  Fluid Mechanics 460 (2002) 211--222.

\bibitem{charrier1989free}
J.~Charrier, S.~Shrivastava, R.~Wu, Free and constrained inflation of elastic
  membranes in relation to thermoforming—non-axisymmetric problems, The
  Journal of Strain Analysis for Engineering Design 24~(2) (1989) 55--74.

\bibitem{helfrich1973elastic}
W.~Helfrich, Elastic properties of lipid bilayers: theory and possible
  experiments, Zeitschrift f{\"u}r Naturforschung c 28~(11-12) (1973) 693--703.

\bibitem{guckenberger2017theory}
A.~Guckenberger, S.~Gekle, Theory and algorithms to compute helfrich bending
  forces: A review, Journal of Physics: Condensed Matter 29~(20) (2017) 203001.

\bibitem{barthes2016motion}
D.~Barthes-Biesel, Motion and deformation of elastic capsules and vesicles in
  flow, Annual Review of fluid mechanics 48 (2016) 25--52.

\bibitem{guckenberger2016bending}
A.~Guckenberger, M.~P. Schraml, P.~G. Chen, M.~Leonetti, S.~Gekle, On the
  bending algorithms for soft objects in flows, Computer Physics Communications
  207 (2016) 1--23.

\bibitem{chorin1968numerical}
A.~J. Chorin, Numerical solution of the navier-stokes equations, Mathematics of
  computation 22~(104) (1968) 745--762.

\bibitem{bell1989second}
J.~B. Bell, P.~Colella, H.~M. Glaz, A second-order projection method for the
  incompressible navier-stokes equations, Journal of Computational Physics
  85~(2) (1989) 257--283.

\bibitem{van2018towards}
J.~A. Van~Hooft, S.~Popinet, C.~C. Van~Heerwaarden, S.~J. Van~der Linden, S.~R.
  de~Roode, B.~J. Van~de Wiel, Towards adaptive grids for atmospheric
  boundary-layer simulations, Boundary-layer meteorology 167~(3) (2018)
  421--443.

\bibitem{johansen1998}
H.~Johansen, P.~Colella, A cartesian grid embedded boundary method for
  poisson's equation on irregular domains, Journal of Computational Physics
  147~(1) (1998) 60--85.

\bibitem{colella2006cartesian}
P.~Colella, D.~T. Graves, B.~J. Keen, D.~Modiano, A cartesian grid embedded
  boundary method for hyperbolic conservation laws, Journal of Computational
  Physics 211~(1) (2006) 347--366.

\bibitem{embed-small-cfl}
S.~Popinet,
  \url{http://basilisk.fr/src/embed.h#lifting-the-small-cell-cfl-restriction},
  accessed: 2022-10-05 (2018).

\bibitem{ghigo-embed}
A.~Ghigo, \url{http://basilisk.fr/sandbox/ghigo/src/myembed.h}, accessed:
  22-11-2022 (2021).

\bibitem{unverdi1992front}
S.~O. Unverdi, G.~Tryggvason, A front-tracking method for viscous,
  incompressible, multi-fluid flows, Journal of computational physics 100~(1)
  (1992) 25--37.

\bibitem{Peskin1977}
C.~Peskin, {Numerical analysis of blood flow in the heart}, Journal of
  Computational Physics 25~(3) (1977) 220--252.

\bibitem{lu2019scalable}
L.~Lu, M.~J. Morse, A.~Rahimian, G.~Stadler, D.~Zorin, Scalable simulation of
  realistic volume fraction red blood cell flows through vascular networks, in:
  Proceedings of the International Conference for High Performance Computing,
  Networking, Storage and Analysis, 2019, pp. 1--30.

\bibitem{farutin20143d}
A.~Farutin, T.~Biben, C.~Misbah, 3d numerical simulations of vesicle and
  inextensible capsule dynamics, Journal of Computational Physics 275 (2014)
  539--568.

\bibitem{meyer2003discrete}
M.~Meyer, M.~Desbrun, P.~Schr{\"o}der, A.~H. Barr, Discrete
  differential-geometry operators for triangulated 2-manifolds, in:
  Visualization and mathematics III, Springer, 2003, pp. 35--57.

\bibitem{uniaxial_stretch.c}
D.~P. Huet,
  \url{http://basilisk.fr/sandbox/huet/tests/lagrangian\_caps/uniaxial\_stretch.c},
  accessed: 2022-10-05 (2022).

\bibitem{biconcave_curvatures.c}
D.~P. Huet,
  \url{http://basilisk.fr/sandbox/huet/tests/lagrangian\_caps/biconcave_curvatures.c},
  accessed: 2022-10-05 (2022).

\bibitem{koleva2012deformation}
I.~Koleva, H.~Rehage, Deformation and orientation dynamics of polysiloxane
  microcapsules in linear shear flow, Soft Matter 8~(13) (2012) 3681--3693.

\bibitem{walter2010coupling}
J.~Walter, A.-V. Salsac, D.~Barth{\`e}s-Biesel, P.~Le~Tallec, Coupling of
  finite element and boundary integral methods for a capsule in a stokes flow,
  International journal for numerical methods in engineering 83~(7) (2010)
  829--850.

\bibitem{nh_shear.c}
D.~P. Huet,
  \url{http://basilisk.fr/sandbox/huet/tests/lagrangian\_caps/nh_shear.c},
  accessed: 2022-10-05 (2022).

\bibitem{zhu2015motion}
L.~Zhu, L.~Brandt, The motion of a deforming capsule through a corner, Journal
  of Fluid Mechanics 770 (2015) 374--397.

\bibitem{le2010effect}
D.~V. Le, Effect of bending stiffness on the deformation of liquid capsules
  enclosed by thin shells in shear flow, Physical Review E 82~(1) (2010)
  016318.

\bibitem{bending_shear.c}
D.~P. Huet,
  \url{http://basilisk.fr/sandbox/huet/tests/lagrangian\_caps/bending_shear.c},
  accessed: 2022-10-05 (2022).

\bibitem{park2013transient}
S.-Y. Park, P.~Dimitrakopoulos, Transient dynamics of an elastic capsule in a
  microfluidic constriction, Soft matter 9~(37) (2013) 8844--8855.

\bibitem{constricted_channel.c}
D.~P. Huet,
  \url{http://basilisk.fr/sandbox/huet/tests/lagrangian\_caps/constricted_channel.c},
  accessed: 2022-10-05 (2022).

\bibitem{pozrikidis2005resting}
C.~Pozrikidis, Resting shape and spontaneous membrane curvature of red blood
  cells, Mathematical Medicine and Biology: A Journal of the IMA 22~(1) (2005)
  34--52.

\bibitem{pozrikidis2010computational}
C.~Pozrikidis, Computational hydrodynamics of capsules and biological cells.

\bibitem{rbc_shear.c}
D.~P. Huet,
  \url{http://basilisk.fr/sandbox/huet/tests/lagrangian\_caps/rbc_shear.c},
  accessed: 2022-10-05 (2022).

\bibitem{caps_interception.c}
D.~P. Huet,
  \url{http://basilisk.fr/sandbox/huet/tests/lagrangian\_caps/caps_interception.c},
  accessed: 2022-10-05 (2022).

\bibitem{caps_interception_inertia.c}
D.~P. Huet,
  \url{http://basilisk.fr/sandbox/huet/tests/lagrangian\_caps/caps_interception_inertia.c},
  accessed: 2022-10-05 (2022).

\bibitem{helix.c}
D.~P. Huet,
  \url{http://basilisk.fr/sandbox/huet/cases/lagrangian\_caps/helix.c},
  accessed: 2022-10-05 (2022).

\bibitem{selccuk2021fictitious}
C.~Sel{\c{c}}uk, A.~R. Ghigo, S.~Popinet, A.~Wachs, A fictitious domain method
  with distributed lagrange multipliers on adaptive quad/octrees for the direct
  numerical simulation of particle-laden flows, Journal of Computational
  Physics 430 (2021) 109954.

\end{thebibliography}

\end{document}